\documentclass[fleqn,usenatbib]{mnras}
\usepackage{newtxtext,newtxmath}
\usepackage[T1]{fontenc}
\DeclareRobustCommand{\VAN}[3]{#2}
\let\VANthebibliography\thebibliography
\def\thebibliography{\DeclareRobustCommand{\VAN}[3]{##3}\VANthebibliography}
\usepackage{graphicx}	
\usepackage{amsmath}	
\usepackage{wasysym}
\usepackage{lastpage}
\usepackage{verbatim}
\usepackage{multirow}
\usepackage{subfigure}
\usepackage{tablefootnote}
\usepackage{hyperref}
\usepackage{times}
\usepackage{graphicx}
\usepackage{float}
\usepackage{booktabs}
\usepackage{threeparttable}
\usepackage{textcomp} 

\title[Spectropolarimetry of Fairall 9]{Linear Spectropolarimetric Analysis of Fairall 9 with VLT/FORS2\thanks{Based on data collected at Paranal Observatory under programme 0102.B-0743(A).}}
\date{Accepted 2021 July 30. Received 2021 July 9; in original form 2021 May 5}
\author[Jiang et al.]{
Bo-Wei Jiang,$^{1,2}$\thanks{jiangbw@ihep.ac.cn}
Paola Marziani,$^{3}$
\DJ or\dj e Savi\'{c},$^{4,5}$
Elena Shablovinskaya,$^6$
Luka \v{C}. Popovi\'{c},$^{5,7}$
\newauthor
Victor L. Afanasiev,$^6$
Bo\.{z}ena Czerny,$^8$
Jian-Min Wang,$^{1,2,9}$
Ascensi\'{o}n del Olmo,$^{10}$
\newauthor
Mauro D'Onofrio,$^{11}$
Marzena \'{S}niegowska,$^{8,12}$
Paola Mazzei,$^3$\ and 
Swayamtrupta Panda$^{8,12}$
\\
\\
$^{1}$Key Laboratory for Particle Astrophysics, Institute of High Energy Physics, Chinese Academy of Sciences, 19B Yuquan Road, Beijing 100049, China\\
$^{2}$School of Astronomy and Space Science, University of Chinese Academy of Sciences, 19A Yuquan Road, Beijing 100049, China \\
$^{3}$INAF, Osservatorio Astronomico di Padova, Vicolo dell'Osservatorio 5, IT-35122, Padova, Italy\\
$^{4}$Institut d'Astrophysique et de G\'eophysique, Universit\'e de Li\`ege, All\'ee du 6 Ao\^ut 19c, 4000 Li\`ege, Belgium\\
$^{5}$Astronomical Observatory, Volgina 7, 11060 Belgrade, Serbia\\
$^{6}$Special Astrophysical Observatory, Russian Academy of Sciences, Nizhnij Arkhyz, 369167, Russia\\
$^{7}$Department of Astronomy, Faculty of Mathematics, University of Belgrade, Studentski trg 16, 11000 Belgrade, Serbia\\
$^{8}$Center for Theoretical Physics, Polish Academy of Sciences, Al. Lotnik\'{o}w 32/46, 02-668 Warsaw, Poland\\
$^{9}$National Astronomical Observatories of China, Chinese Academy of Sciences, 20A Datun Road, Beijing 100020, China\\
$^{10}$Instituto de Astrofis\'{i}ca de Andaluc\'{i}a, IAA-CSIC, E-18008 Granada, Spain\\
$^{11}$Dipartimento di Fisica \& Astronomia ``Galileo Galilei'', Universit\`{a} di Padova, Padova, Italy\\
$^{12}$Nicolaus Copernicus Astronomical Center (PAN), ul. Bartycka 18, 00-716 Warsaw, Poland
}
\def\hb{H$\beta$\/} 
\def\ha{H$\alpha$\/}
\def\rfe{{$R_{\rm Fe {II}}$}\/}
\def\kms{$\rm km\ s^{-1}$\/}
\def\msun{$\rm M_{\rm \astrosun}$\/}
\def\mbh{$M_{\rm BH}$\/}
\def\edd{$L_{\rm Bol} / L_{\rm Edd}$\/}
\def\ergs{$\rm {{10^{44}\,erg\,s^{-1}}}$\/}
\begin{document}
\label{firstpage}
\pagerange{\pageref{firstpage}--\pageref{lastpage}}
\maketitle
\begin{abstract}
The quasar Main Sequence (MS) appears to be an incredibly powerful tool to organize the diversity in large samples of  type-1 quasars but the most important physical parameters governing it are still unclear.  Here we investigate the origin of the broadening and of a defining feature of Population B sources: a strong redward asymmetry of the Balmer emission lines. We focus  on a prototypical source, Fairall 9.  Spectropolarimetric data of the Fairall 9  broad H$\beta$ and H$\alpha$ profiles allowed for a view of the geometric and dynamical complexity of the line emitting regions. Measurements (1) provided evidence of rotational motion; (2) were helpful to test the presence of polar and equatorial scatterers, and their association with  non-virial motions. 
{However, we  suggest that the polarization properties appear to be more  consistent with a warped disk geometry induced by Lense-Thirring precession. }
\end{abstract}
\section{Introduction}

The quasar Main Sequence (MS) is a concept that arose from the Principal Component Analysis on 87 quasars introduced by \citet{BG92}. 
The usefulness of the MS is rooted in the ability to contextualize every quasar as part of a sequence \citep{Sulentic2000blr,ShenHo2014}. Most notable correlations involve the width and shape of the Balmer line profiles, the strength of optical Fe {\sc ii} emission and the amplitude of the systematic blueshifts of high-ionization lines with respect to the quasars rest frame \citep[see e.g.,][for a summary]{fraix-burnetetal17}. Eventually,  the { Eigenvector 1 (E1)}-related correlations  allowed for  the identification of two main quasar populations along the MS: Population A (Pop. A) with FWHM(\hb) \ensuremath{\le} 4000 \kms\, and Population B (Pop. B) with FWHM(\hb) > 4000 \kms\ \citep{Sulentic2000blr}.

The physical parameters governing the MS are still being investigated, although some basic inferences have been made.  
The main parameters that describe quasars as accreting black holes  are the black hole mass \citep[\mbh, ranging from $10^6$\msun\ to $10^{9.5}$\msun;][]{KR95}, the accretion luminosity, Eddington ratio \citep[\edd,][]{BG92,Sulentic2000e1,PM2001,PM2003MN}, and the black hole spin \citep[e.g.,][]{wangetal14}. The \mbh\ can be estimated by assuming that the gas motions are predominantly Keplerian around the black hole, and by applying the virial theorem for a system whose mass is entirely concentrated in the center of gravity. Reverberation mapping provides a measurement of the radial distance $r_\mathrm{BLR}$\ of the line emitting gas from the central black hole \citep{peterson93,peterson2004}. The so-called "virial mass" have been computed for large samples of quasars employing several different  emission lines, various measures of line width, and exploiting correlation between the emitting region radius and luminosity \citep[see ][for  reviews]{marzianisulentic12,shen13,pop20}. An independent method relies on the scaling law between \mbh\ and the stellar velocity dispersion of the galaxy bulge \citep[$\sigma_\star$,][]{KHo2013}. Recently, spectropolarimetric observations of the Hydrogen Balmer lines have allowed to measure the intrinsic line FWHM due to a Keplerian velocity field and  to compute the black hole mass independently from orientation {\citep[][and references therein]{Afa2015,Afa2019}}. All methods to compute black hole mass in AGN are subject to caveats and suffer considerable uncertainties \citep[e.g., ][and references therein]{dallabontaetal20}.   Nonetheless, the \edd, which is proportional to the luminosity-to-black hole mass ratio ($L/$\mbh), has been revealed to be a fundamental driver of the E1 MS, closely related to several observational parameters \citep{PM2001,PM2003MN,kuraszkiewiczetal04,ShenHo2014,pandaetal18,pandaetal19}.  With decreasing Eddington ratio,  source properties change from the ones of Pop. A to the ones of Pop. B, which tend to show broader Balmer emission lines, weaker \rfe\footnote{Defined as the ratio of the intensity of Fe {\sc ii}$\lambda4570$ blend to the intensity of \hb\ broad component}, and more asymmetric Balmer line profiles \citep{Sulentic2000e1,PM2003MN,ShenHo2014}.

Apart from the Eddington ratio, several studies found  that the orientation effects can influence the FWHM of Balmer lines in type-1 quasars as well  \citep[e.g.,][]{Wills86,rokakietal03,Sulentic2003,jarvismclure06,decarlietal11,pandaetal19}. 
In this sense, Pop. B sources might be seen at a larger viewing angle \citep[defined as the angle between the line-of-sight and the accretion disk axis;][]{PM2001,ShenHo2014}. Other factors influence the line widths in addition to orientation, most notably \mbh, and \edd. From the position of a source in the optical plane of the MS, FWHM(\hb) vs \rfe{}, it is not possible to retrieve unambiguous evidence on the viewing angle. This is an unfortunate occurrence, as there is observational and theoretical support for a highly-flattened low-ionization emitting region \citep[e.g.,][and references therein]{mejia-restrepoetal18}.  

In addition to broader line profiles, Pop. B sources tend to have  red-ward asymmetries in Balmer emission lines (see e.g., \citealt{BG92,PM96,sulenticetal02,punsly10,wolfetal20}). Several studies have found out that the \hb\ emission line profiles can be empirically modeled by   a very-broad component (VBC) with a typical width (\ensuremath{\gtrsim} 10,000 \kms), twice as broad as the classical broad component\citep[BC,][]{Sulentic2000blr}. The VBC shift is yielding  the observed asymmetry \citep{PM2003}. The physical explanation of the unshifted BC involves a virialized, optically-thick gaseous region \citep{SG2007,czerny2011,wang2017,PM2018}. The VBC can be explained as due to a high-ionization, at least in part virialized region  closer to the central black hole \citep[called the very broad line region, VBLR;][]{Peterson86,Mike94,pop95,Sulentic2000blr,pop04,SG2007,WangLi2011,PM2019}. 

The projection effect due to different viewing angles of broad line region (BLR) and the physical origin of the red-ward asymmetry are  still enigmatic at the time of writing.  Spectropolarimetric studies provide a powerful tool for probing the geometry of the BLR \citep{Smith2005,Afa2014,Baldi2016,Afa2019}. The polarized light, scattered from an equatorial dusty torus, contains structural and kinematic information for both the BLR and the scattering region \citep{Smith2005}. The measurement of the Stokes parameters, the polarization degree ({$P\%$}), the polarized flux ({$P\times F$}) and the polarization position angle ({PA})  can in principle be used for estimating the mass of the central SMBHs in a way that is independent of the viewing angle \citep{Afa2015,Savic2018,songsheng2018,Afa2019,savic2020},  as a potential tracer of super-massive binary black holes as well as of non-virial motion\citep{Savic2019} .

Fairall 9 \citep[$\equiv$ ESO 113--45;][]{Fairall} is a  Seyfert 1 galaxy at a comoving radial distance of $\approx 195.3$  Mpc, with a luminous  nucleus \citep[$M_V$ = -22.6;][]{veron2000}.  Optical spectroscopy in the \hb\ spectral range indicates that Fairall 9 is a  prototypical Pop. B quasar, with FWHM of the {virialized} \hb\ BC at 4550 \kms\ and \rfe\ $\lesssim 0.5$\ \citep{PM2010}. Moreover, it has a moderate redward asymmetry that can be modeled with  a VBC shifted to the red in Ly$\alpha$, C {\sc iv}, Mg {\sc ii} and \hb\ emission lines \citep{PM2010}. 

 In this paper we present analysis of the new spectropolarimetric observations of Fairall 9 AGN with the aim to investigate the innermost part and  to ascertain the origin of the broadening and of the redward asymmetry of the \ha\ profile { observed in this object.}  More details on the source are reported in Section  \ref{f9}. We then present the analysis of spectropolarimetric data  obtained using the Very Large Telescope (VLT, Section  \ref{obsredpol}).   Main results from  parameters measured on the polarization spectrum are  described in Section  \ref{results}. Results point to  a more complex scenario than    a single-disk Keplerian motion and an equatorial scattering. They are discussed and interpreted with the help of ray-tracing programs in Section  \ref{disc}. { Throughout the paper, we adopt a ${\Lambda}$CDM cosmological model with $H_0=70$ km s$^{-1}$ Mpc$^{-1}$, $\Omega_\mathrm{M}$=0.3, and $\Omega_{\Lambda}=0.7$.}


\section{Fairall 9: a prototypical Population B source}
\label{f9}

Since the time of its identification, Fairall 9 has been considered a favored target because of its brightness in the IR, optical, UV and X-ray domains. At first, it was so because Fairall 9 was believed to be an "extreme" Seyfert 1 where extreme meant that it was bordering the luminosity range of quasars.  The host galaxy is well-resolved, classified as S0 by the APM Bright Galaxy Catalogue \citep{loveday96}, more likely SB0a on a visual inspection of the ACS image shown by \citet{bentz2009}, with a faint companion (LEDA 5109) located at $30$ arcsec SSW from Fairall 9. The angular separation corresponds to $\sim 30$\ kpc of projected linear distance.  Fairall 9 is definitely radio quiet, as it was not detected by the Sydney University Molonglo Sky Survey (SUMSS, \citealt{mauchetal03}) with a detection limit of 6mJy, implying a ratio radio-to-optical specific flux $\lesssim 1.5$.  

 Reverberation mapping campaigns in the optical \citep{santos-lleoetal97} and in the UV \citep{VP2006} have provided measurements of the BLR radius $r_\mathrm{BLR}$\ from both low- and high-ionization lines.  Previous  estimates of black hole  mass based on the reverberation-mapping method converge to values  $(1.5 - 2.5)\times10^{8}{\rm M_{\astrosun}}$ \, \citep{peterson2004,bentzkatz15}, depending on the adopted virial factor. { A conventional estimate of the bolometric  luminosity for Fairall 9 is $\log L_{\rm Bol} \approx 45.3$\ erg s$^{-1}$, and the Eddington ratio $\log$(\edd) $\approx -2.0$\ \citep{PM2010}, close to the low end in the distribution of  Pop. B sources \citep{PM2003MN}.} { Revised estimates of these parameters are provided in Section  \ref{sect-mass}.  }

Line shifts with respect to rest-frame are of special importance to this investigation. The rest frame we assumed is based on the measured redshift of Fairall 9,  $z = 0.04609 \pm $0.00002 (heliocentric redshift is $z_{\odot} \approx 0.04605$) measured on a 1993 ESO spectrum published in \citet{PM2003}, from the narrow component of \hb\ and from [O{\sc iii}]$\lambda\lambda$4959,5007).\footnote{The 10 \kms\ heliocentric correction is not relevant considering the scale of the polarization spectrum is $\approx$ 4 \AA/pix $\approx$ 180 \kms).}

\begin{figure}
\centering
\includegraphics[width=\linewidth]{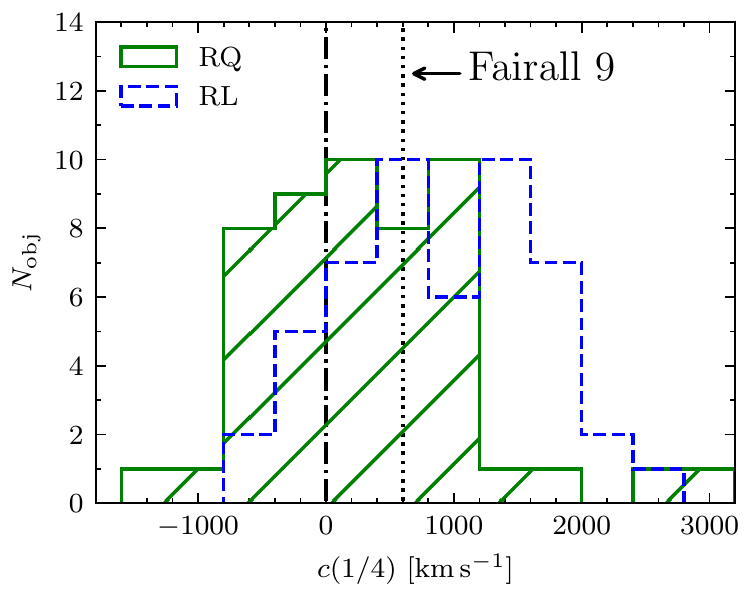}
\caption{\label{c1_4}The measured \hb\ centroid at 1/4 of the maximum, based on the sample of \citet{PM2003} of 101 Pop.B objects. The green histogram refers to the distribution of 51 radio-quiet (RQ) objects, while the blue histogram represents the 50 radio-loud (RL) sources. { The $c(1/4)$ of Fairall 9 is  $\approx 600$ \kms\ marked as dotted line in the plot.} The 0-velocity shift is marked as dotted-dash line. { The bin size is 400 \kms\ for both RQ and RL sources, and was set on the basis of the typical uncertainty in the $c(1/4)$ estimates.}}
\end{figure}

 The MS empirical parameters reveal that the  spectral type of Fairall 9 is B1, one of the most populated along the MS of quasars \citep{Sulentic2000e1,PM2003MN,ShenHo2014}, consistently associated with low \edd. { Following \citet{zamfiretal10}, we define the centroid shift { with respect to the rest-frame} at 1/4 maximum  $c(1/4)=(v_{\rm B}(1/4)+v_{\rm R}(1/4))/2$, where $v_{\rm B}(1/4)$ and $v_{\rm R}(1/4)$ refer to the velocity shift on the blue and red wing at 1/4 of the peak intensity, respectively. { The $c(1/4)$ is a measurement of the line profile displacement with respect to the rest frame and its value measured on the \hb\ profile of Fairall 9 is  $\approx$ 600 \kms\ } close to the average and median values for the Pop. B RQ sample  of \citet[][shown in Fig.   \ref{c1_4}]{PM2003}\footnote{\url{http://vizier.u-strasbg.fr/viz-bin/VizieR?-source=J/ApJS/145/199}}}.  { The spectral energy distribution (SED)} is also consistent with objects of Pop. B with no strong big-blue bump.  {In the X-ray domain, Fairall 9 has a flat spectrum { ($\Gamma = 2.0$)}, with  photon index from the 0.5 to 40 keV  $\Gamma = 1.8-2.0$  \citep{Loh2016}. The X-ray spectrum has been interpreted as due to a power law,  continuum components ({ including a modest soft X-ray excess}) associated with cold and ionized blurred reflection \citep{Emm2011,Walton2013}.}

\section{Observations, Data Reduction, and Polarization Analysis}
\label{obsredpol}

\subsection{Observations}

The FOcal Reducer and low dispersion Spectrograph 2 (FORS2) is  installed at the Cassegrain focus of ESO's Very Large Telescope (VLT). The spectrograph has been designed for multiple purposes in visible and near ultraviolet bands, including a polarimetric mode.
The polarimetric mode allows the measurement of linear polarization. A Wollaston prism is introduced as beam splitter in the optical path, and a super-achromatic half-wave retarder plate connected to a grism with slitlets and a filter make it possible to carry out spectropolarimetry \citep{ESO}.

For our observations, we used the GRIS\_300V grism and the order separation filter GG435 to fully cover the  { \hb\ and}  \ha\ profile, and approaching a spectral resolution $\sim 800$ in the wavelength range of \ha, with a step of $\approx$ 4 \AA/pix. The observation of
Fairall 9 was carried out during the night of November 7, 2018, with seeing condition $\apprle1$ arcsec. The half-wave retarder plate was rotated at angles of $0^{\circ},\ 22.5^{\circ},\ 45,\ {\rm and}\ 67.5^{\circ}$. At every individual position angle, a set of six observations was taken, each with an exposure of 240 seconds. The centroid of the target was placed in one of the Multi-Object Spectroscopy movable slitlets
(MOS), with the height set to 11.4 arcseconds. The produced spectra were split into ordinary (o) and extraordinary (e) beams by the Wollaston prism, and recorded on the CCD with a separation of about 22 arcseconds (the pixel scale in the spatial direction is $D_{{\rm s}}=0.126\ {\rm arcsec\,/\,pixel}$\ \citep{ESO}.


\subsection{Data Reduction}

 { We performed the standard procedure of the spectropolarimetric data reduction described in details in \citet{afa2012} and adopted to the VLT/FORS2 data. The reduction process includes the following steps: bias subtraction and extraction of the frames from the original fits-files, construction of the 2D model of the geometrical distortions for ordinary and extra-ordinary ray spectra, spectra linearization using He-Hg-Cd-Ar arc lamp spectrum and flat-field correction and night sky lines subtraction. The effective spectral range is 4400--7500 \AA.

The o- and e-beam spectra were extracted using a fix aperture size of around 40 pixels. The size of the aperture was chosen so that the nucleus flux was totally integrated within the seeing on each exposure. 
\subsection{\label{subsec:Polarization-analysis}Polarization analysis}

The formul\ae\ to calculate the normalized Stokes parameters in the general form can be found in \citet{PR2006}. Here we will use the relations adopted for the case of the measuring of the polarization with the Wollaston prism and rotating $\lambda$/2 plate given in \citet{afa2012}:
\begin{equation} \label{qu1}
    \begin{split}
        Q(\lambda) &= \frac{1}{2}  \left( F(\lambda)_{\theta=0^\circ } -  F(\lambda)_{\theta=45^\circ}  \right), \\
        U(\lambda) &= \frac{1}{2}  \left( F(\lambda)_{\theta=22.5^\circ } -  F(\lambda)_{\theta=67.5^\circ}  \right), \\
    \end{split}
\end{equation}
where $\theta$ is the position angle of the retarder plate and $F(\lambda)$ is the normalized flux difference between the ordinary ($f_{o}(\lambda)$) and extraordinary ($f_{e}(\lambda)$) beams:
\begin{equation}
F(\lambda)=\frac{f_{o}(\lambda)-f_{e}(\lambda)}{f_{o}(\lambda)+f_{e}(\lambda)}\label{fofe}.
\end{equation}
}
Linear polarization degree $P$ and polarization position  angle are related to the Stokes parameters as:
\begin{equation} \label{qu}
    \begin{split}
        P(\lambda) &= \sqrt{Q(\lambda)^{2}+U(\lambda)^{2}}, \\
        \mathrm{PA}(\lambda) &= \frac{1}{2}\arctan\frac{U(\lambda)}{Q(\lambda)}. \\
    \end{split}
\end{equation}
The $\pi/2$ ambiguity of the polarization angle is corrected according to the formulae given in \cite{Bagnulo2009}.


\begin{table*}
\centering

\caption{{ \label{tab:ism} Low-polarized star properties}}
\setlength{\tabcolsep}{3.3\tabcolsep}
\begin{tabular}{ccccccc}
\toprule 
RA & DEC & $P$ & PA & $Q$ & $U$\tabularnewline
(h:m:s) & (d:m:s) & (\%) & ($^\circ$) & (\%) & (\%)\tabularnewline
(1) & (2) & (3) & (4) & (5) &(6) \\
\midrule
01:06:54.46 & -59:58:01.2 & 0.016$\pm$0.023 & 110.0$\pm$35.7 & -0.012$\pm$0.022 & -0.010$\pm$0.021 \tabularnewline
01:31:32.58 & -59:35:34.4 & 0.012$\pm$0.014 & 83.9$\pm$30.3 & -0.012$\pm$0.014 & 0.003$\pm$0.013 \tabularnewline
01:35:14.71 & -58:08:21.5 & 0.015$\pm$0.012 & 97.1$\pm$21.8 & -0.015$\pm$0.012 & -0.004$\pm$0.011\tabularnewline
\bottomrule
\end{tabular}

\begin{minipage}{\linewidth}
\vspace{0.2cm}

{ \textit{Notes}: The three low-polarized stars from \citet{Heiles2000}. Col. (1), right ascension. Col. (2), declination. Col. (3) and (4), polarization degree($P$) and polarization position angle(PA) reported in the literature. Col. (5) and (6), the decomposed Stokes parameters $Q$ and $U$ computed from Col. (3) and (4) based on Eq. \ref{qu}. All the uncertainties reported here are at $1\sigma$ confidence level}
\end{minipage}
\end{table*}

\begin{figure}
    \centering
    \includegraphics{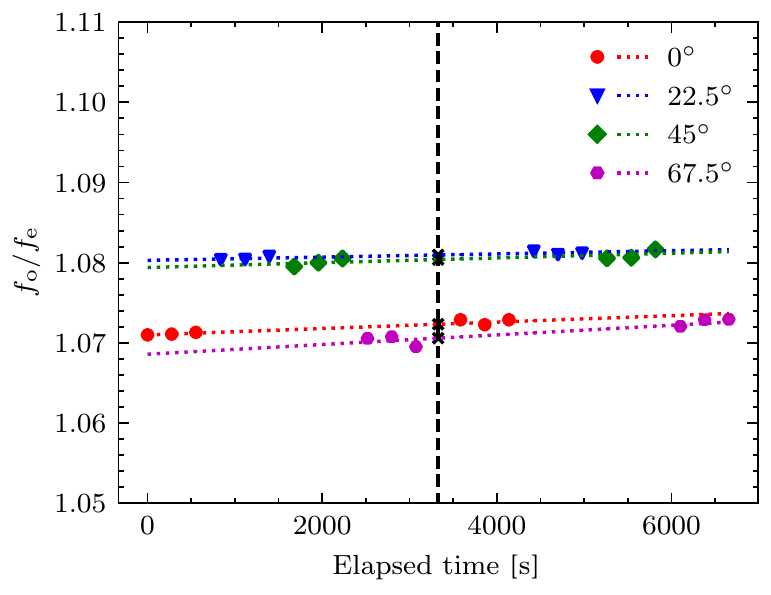}
    \caption{The change of depolarization coefficients in polarizing channels during observations for different retarder angles. The scatters are the average ratio in each frame. The dotted lines show the tendency of the ratio. Black dashed line is at the middle of observation with cross mark showing the speculative depolarization coefficients. }
    \label{fig:depol}
\end{figure}

\subsubsection{Atmospheric depolarization}
The atmospheric depolarization, caused by non-selective aerosol light scattering, has a serious effect on the results of the spectropolarimetric analysis. From the data we collected, the effect could reach up to 1\%  (shown in Fig.  \ref{fig:depol}). To eliminate such effect, we followed procedures discussed in \citet{afa2012}. We assumed the depolarization effect is wavelength independent, and we considered the  depolarization during the exposure as a variation of the spectrograph transmission for the ordinary and extraordinary rays. In this case, we used 
\begin{equation}
    F(\lambda)=\frac{D(\theta_i)f_{o}(\lambda)-f_{e}(\lambda)}{D(\theta_i)f_{o}(\lambda)+f_{e}(\lambda)}\label{fofea}
\end{equation}
instead of the expression in Eq. \ref{fofe}, where $D(\theta_i)$ are the coefficients of the polarization transmission channels which also include the variations of the atmospheric depolarization. As it can be found in \citet{afa2012} these coefficients are usually obtained for each retarder position angle and for each exposure. Due to the non-optimal observational technique we had to use the time-averaged coefficients which are calculated as:
\begin{equation}
D(\theta_i) = \langle {f_{\rm o}(\lambda)}/{f_{\rm e}(\lambda)} \rangle
\end{equation}

where the angle brackets correspond to the averaging over the time.
Correction because of atmospheric (de)polarization should be  applied every time the ratio between the ordinary and extraordinary beam intensity is shown to depend on time. Even a change of $\approx$1\%\ can yield to a large difference in the polarization angle.

\subsubsection{Interstellar polarization}

The interstellar matter (ISM), including molecular gas and dust, can
have a considerable effect on the linear polarization because of scattering
\citep{ism2001}. The observed polarization is a vectorial composition
of the polarization of the target and ISM, {and the corresponding Stokes parameters $U$ and $Q$ have the relation as: } 
\begin{equation}
    {(Q,U)}_{{\rm obs}}={(Q,U)}_{{\rm AGN}}+{(Q,U)}_{{\rm ISM}}
\end{equation}

To determine and eliminate the ISM polarization, we used 3 low-polarized stars \citep{Heiles2000} close to Fairall 9. Their polarization data are  in Table  \ref{tab:ism}. We assume that all 3 stars are non-polarized, and that their observed polarization is entirely caused by ISM scattering. Even assuming this one can clearly see that the polarization of the nearby stars is equal to zero within the very low errors. So, we will consider further that the interstellar polarization is negligible in the direction of Fairall 9.

\begin{figure*}
\centering

\includegraphics[scale=1]{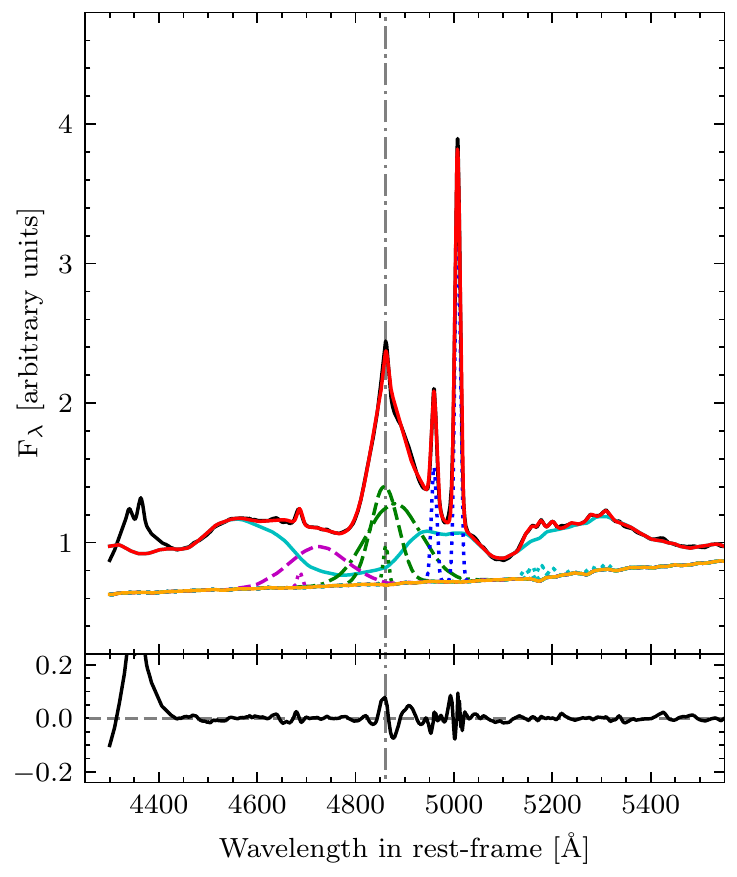}
\includegraphics[scale=1]{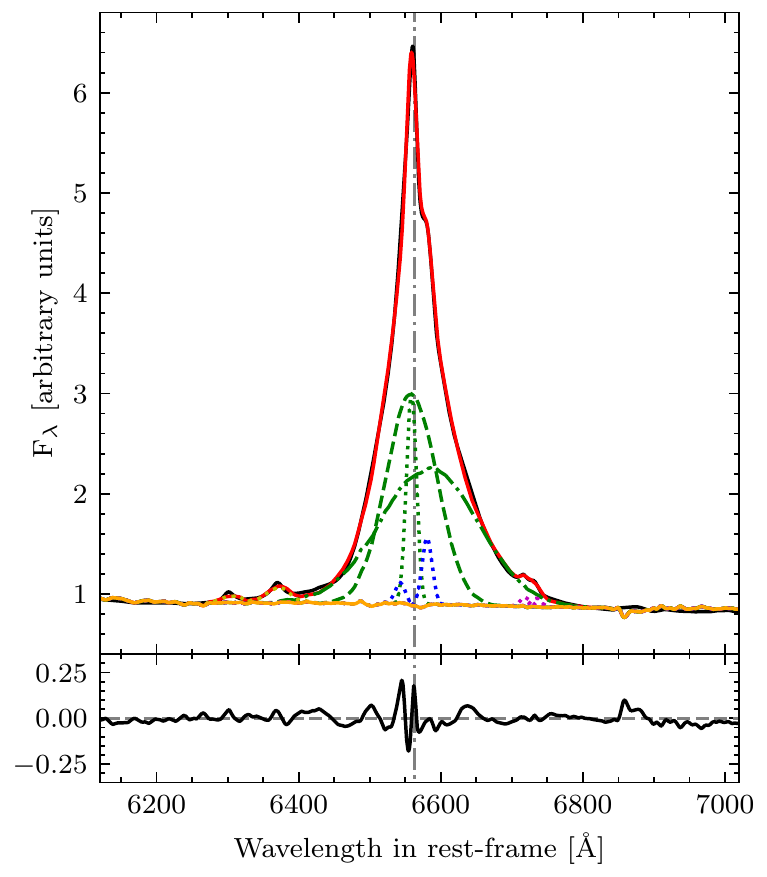}
\caption{\label{fig:fit}{Decomposition of the \hb\ (left panels) and \ha\ (right panels) natural light profiles with multiple components. The grey dot-dashed line marked the rest frame for \hb\ and \ha\ respectively. Top-left panel: the fitting range for \hb,  4430--5550 \AA. The observed total flux is in solid black line while the best-fitting result is shown in red. The sum of the power-law continuum and the host galaxy component is shown as an orange line. The Fe {\sc ii} blends modeled with the template are plotted in solid cyan line. Green dashed, dot-dashed and dotted lines represent the decomposed BC, VBC and NC for \hb, respectively.  Blue dotted lines are {[}O {\sc iii}{]}$\lambda\lambda$4959, 5007. He {\sc ii}$\lambda4686$ emission are modeled with a VBC (magenta dashed line) and a NC (magenta dotted line). Dotted cyan lines stand for the high-ionization iron forbidden lines in the range 5100 -- 5300 \AA. Top-right panel: the fitting range for \ha, 6200--7000 \AA. The black, red and orange solid lines have identical meaning as in left panel. Green dashed, dot-dashed and dotted lines represent the decomposed BC, VBC and NC for \ha, respectively. Blue dotted lines are {[}N {\sc ii}{]}$\lambda\lambda$6548, 6584. Yellow dotted lines are [O {\sc i}]$\lambda$6300 and [O {\sc i}]$\lambda$6363. Magenta dotted lines represent [S {\sc ii}]$\lambda\lambda$6717,6730. Bottom panels are the residual for the fitting with grey dashed lines marking the zero levels.}}
\end{figure*}

\begin{figure}
    \centering
    \includegraphics[scale=1]{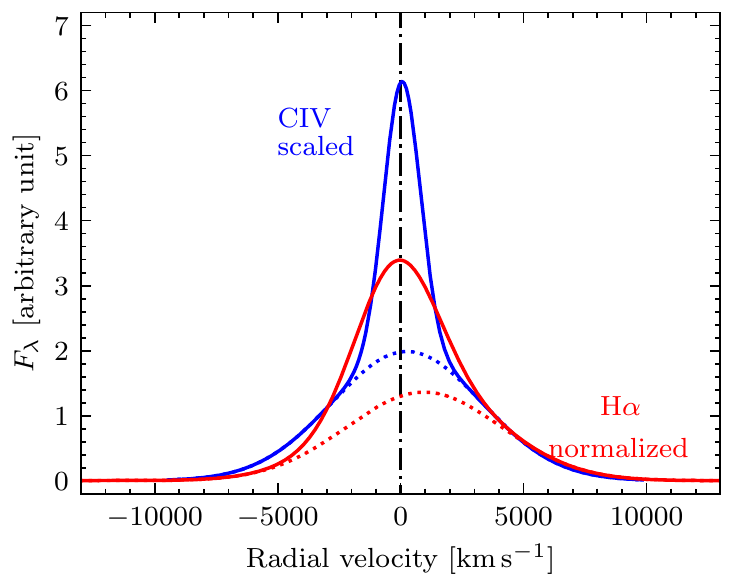}
    \caption{\label{fig:civ0} \ha\ (red) and C {\sc iv}$\lambda$1549 (blue) profiles overlaid after rescaling of C {\sc iv}$\lambda$1549 to the same intensity of the \ha\ red wing. \ha\ and C {\sc iv}$\lambda$1549 VBC are shown by dotted red and blue lines, respectively.}
    
\end{figure}

\subsubsection{Instrumental polarization}

To correct the bias of the mean level of the Stokes parameters calculating from the given spectropolarimetric data we need to correct the data for the instrumental polarization.  { Because no standard star of zero polarization was obtained at the night of Fairall 9 observations, the zero level of the instrumental polarization was defined from the spectra of the host-galaxy of the object. For this, we integrated the frames of the spectra in the range out of the object aperture. Assuming the instrumental polarization independent from the wavelength we obtained: $Q_{\rm ins} = 0.8,\ U_{\rm ins} = -0.1$. The instrumental values were subtracted from the Stokes parameters of the object.}

\section{Results}
\label{results}

\subsection{Natural light decomposition}
We first decomposed the \ha\ and \hb\ emission line profile in  natural light { (shown in Fig.  \ref{fig:fit})}. A minimum-$\chi^2$\ analysis using the IRAF task {\tt specfit} \citep{kriss94} was carried out to include all the relevant components in the \ha\ spectral region from 6200 to 7000\AA\  with { a power-law AGN continuum}, host galaxy continuum, \ha\ emission components, narrow forbidden lines of [O {\sc i}]$\lambda$6300, [O {\sc i}]$\lambda$6363, [N {\sc ii}]$\lambda\lambda6548,6584$,  and [S {\sc ii}]$\lambda\lambda$6717,6730. 
As for \hb, the fitting was carried out in the spectral range from 4430 to 5550\AA, including { a power-law AGN continuum, host galaxy continuum, \hb\ emission components, the Fe {\sc ii} emission blends, the narrow forbidden lines [O {\sc iii}]$\lambda\lambda4959,5007$, He {\sc ii}$\lambda4686$ emission, and  Gaussians accounting for the blended emission of several  high-ionization iron forbidden lines peaking at $\lambda \approx 5200$ \AA\  ([Fe {\sc vi}]$\lambda5146$, [Fe {\sc vii}]$\lambda5159$, [Fe {\sc vi}]$\lambda5176$, [Fe {\sc iv}]$\lambda5236$, [Fe {\sc vii}]$\lambda5276$, and [Fe {\sc xiv}]$\lambda5302$}).
{ Since the host contamination of Fairall 9 is relatively weak with respect to the active nucleus continuum, the host galaxy component in natural light has been modelled  using the spectrum of the elliptical galaxy NGC 3379 \citep{kennicutt92}. The Fe {\sc ii} emission templates are based on I Zw 1 from \citet{BG92}.} 
The decomposition of the \ha\ profile proper involved 3 Gaussians (Fig.   \ref{fig:fit}): (1) the narrow component (NC), (2) the broad component ({BC}), (3) the very broad component (VBC) which has a substantial peak shift to the red by $\thicksim\ 1000$ \kms, and accounts for the redward asymmetry of the \ha\ full broad profile (i.e., BC+VBC). 
 { The decomposition of \hb\ also involved 3 Gaussians (see also Fig.  \ref{fig:fit}): (1) one NC, (2) one BC, (3) one VBC that is shifted to the red by $\thicksim 2000$\kms,  leading to a more expanded red wing in the total \hb\ profile}.

In radio-quiet Pop.B sources, C {\sc iv}$\lambda1549$ shows a systematically stronger blue side with respect to the Balmer line profiles. The excess is small and much lower than in several cases of extreme Pop. A sources that have blueshifted emission dominating the C {\sc iv}$\lambda1549$\ profile \citep[e.g.,][for examples]{leighlymoore04,marinelloetal20b}. The origin of the blue excess in natural light is often considered to be due to outflowing gas, possibly in the form of a wind, dense clumps, or dense clumps embedded in a wind (see e.g. \citealt{kollatschny03,proga03}). 

{ A HST/FOS spectrum was extracted from the MAST archive and reduced by \citet{sulenticetal07} who provide measurements  of the C{\sc iv} line intensity and profile parameters. } The C {\sc iv} $\lambda1549$\ profile  of Fairall 9  has been reanalyzed in this work, with special attention to the line wings. We carried out the decomposition of the C {\sc iv} $\lambda1549$ based on HST/FOS UV spectra following the approach developed in \citet{PM2010}.  We consistently used the same components in the fitting procedure, i.e. one NC, one BC and one VBC. The result is shown in Fig.   \ref{fig:civ0}. Compared to \ha, C {\sc iv} seems to be more symmetric, with a redshifted VBC at only a few hundred \kms. This is mainly caused by two unresolved Gaussian components shifted to blue and red respectively \citep{PM2010}. Since the redshifted component is slightly more prominent, we can see a modest redward asymmetry in total flux.  Now to help the interpretation of the \ha\ profile, we superimpose and rescale the red-wing of C {\sc iv} and \ha\ at the same intensity (Fig.   \ref{fig:civ0}). Both \ha\ and C {\sc iv} have   outstanding redward asymmetries. { The red side of C {\sc iv} is   consistent with the one of \ha,} but with a revealing difference: the centroid at 1/4 peak intensity $c(1/4)$ is 194 $\pm$199 \kms\ for C {\sc iv} and 442$\pm$	136 \kms\ for \ha, where the uncertainties are at a $2\sigma$ confidence level. If the red wings are matched, the blue side of C {\sc iv} is somewhat more extended, making the C {\sc iv} profile   more symmetric especially close to the line base: { the shift to the red of the $c(1/4)$\ is  only marginally significant.} This implies that the C {\sc iv} is affected by a small blueshifted excess, most likely associated with  outflowing gas. 



\subsection{Polarization of the BLR}
\label{polblr}

\begin{figure*}
\centering
\includegraphics[scale=0.8]{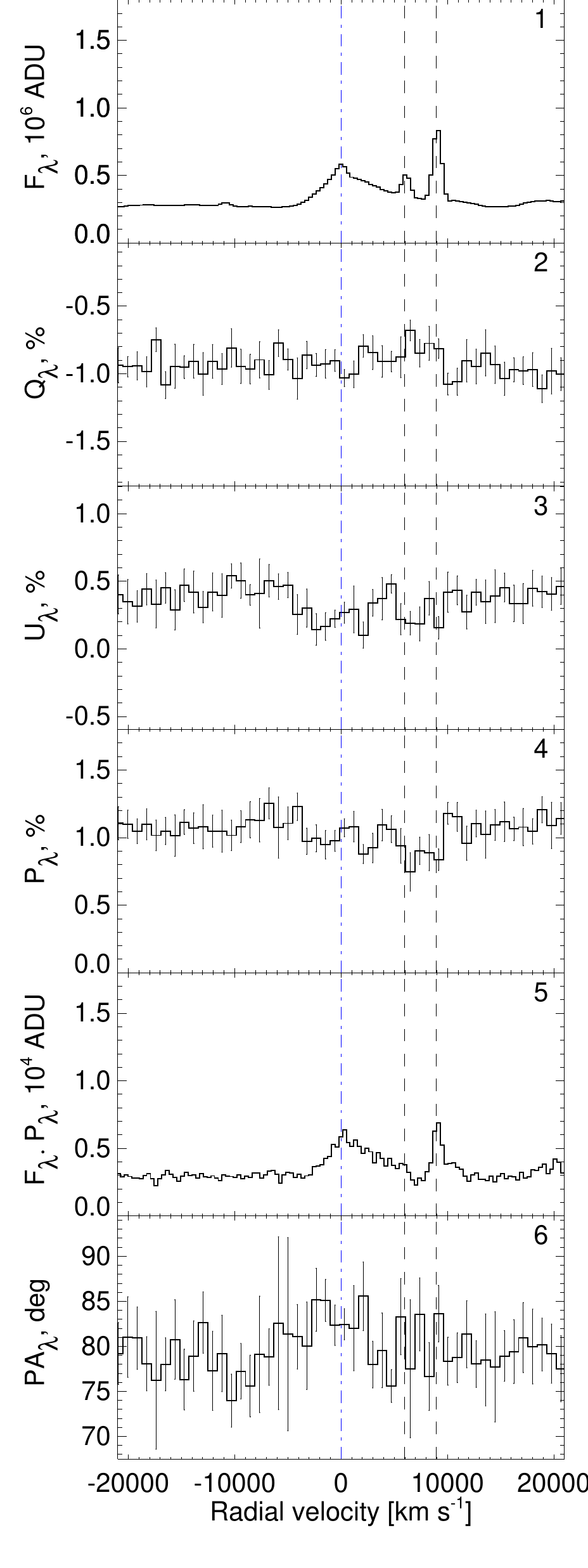}
\includegraphics[scale=0.8]{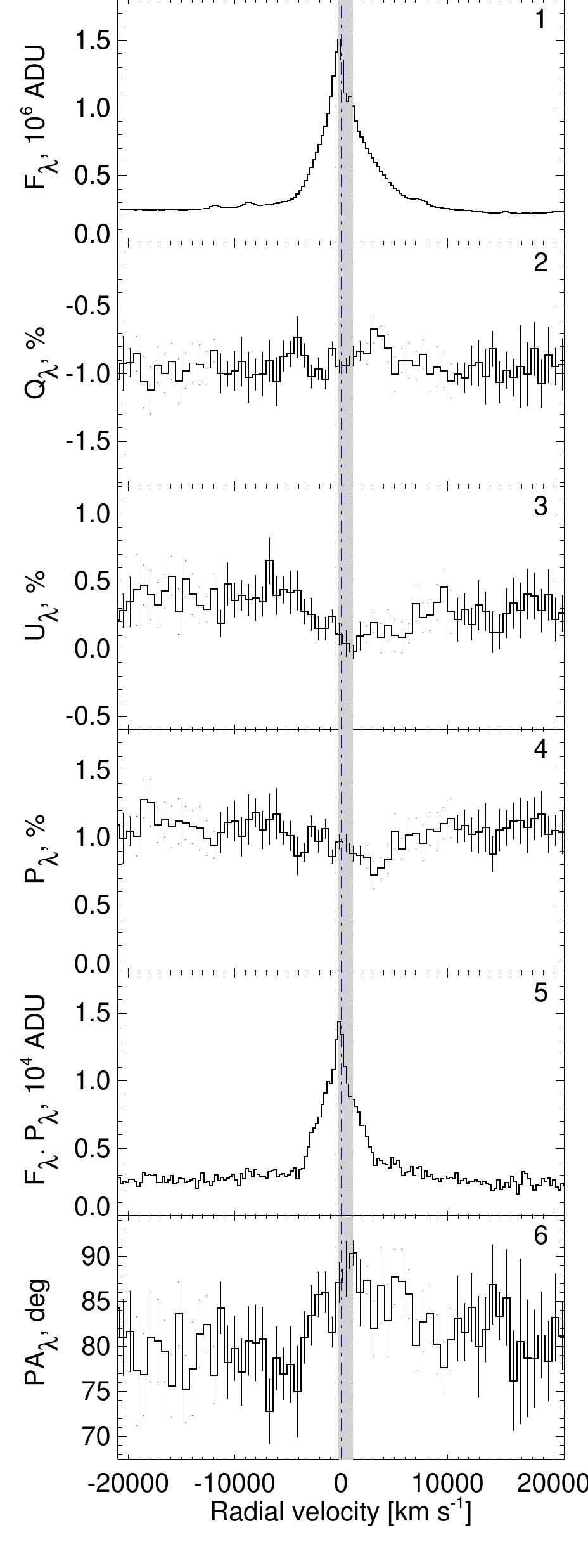}
\caption{\label{fig:stokes}{The profiles of \hb\ (left) and \ha\ (right) lines. From top to bottom: the intensity in the natural light, the Q and U Stokes parameters, the polarization degree, the polarized flux and the polarization angle. The values on panels 1 and 5 are binned in 6\AA\ window and the values on panels 2--4 and 6 -- in 15\AA\ window. The black vertical dashed lines mark the narrow lines coming from the NLR -- [O {\sc iii}] lines at \hb\ profile and [N {\sc ii}] lines at \ha\ profile. The blue vertical dashed line marks the position of zero velocity. For \ha\, the atmospheric absorption band is noted with a light grey stripe. { Error bars are at  1$\sigma$\ confidence level. } }}
\end{figure*}

After the correction for the atmospheric depolarization, ISM and instrumental polarization, we  derived the polarization parameters using the method described in  \ref{subsec:Polarization-analysis}. The results are presented in Fig.  \ref{fig:stokes}. 

 { The total intensity profiles of  \hb\ and \ha\ lines are shown in  panel 1 of Fig.  \ref{fig:stokes}.}  { As the contribution of the narrow lines ([O {\sc iii}] and [N {\sc ii}]) will depolarize the broad lines slightly but their wrong estimation will cause much larger inaccuracies, the narrow components were not subtracted from the spectra. Note here that the atmospheric absorption B-band affects  the profile of \ha\ line between 6860--6917 \AA.}  

 { The 2nd and 3rd panels in Fig.  \ref{fig:stokes} show the Stokes parameters $Q$ and $U$ binned over the 15\AA-window. For each spectral bin, we estimated the robust average of the measured values over the spectral range and all taken exposures. The errors given on the plots are equal to the 1$\sigma$ level, where $\sigma$ is the robust standard deviation\footnote{Note here that we are applying the basic robust estimation using 2$\sigma$ rejection threshold to avoid the influence of the outlier points in the observational data set. The algorithm implementation could be found in more details in \citet{numres}, and references therein}. }

The polarization degree $P$ is plotted at the 4th panel in Fig.  \ref{fig:stokes}.  { The \ha\ and \hb\ lines   show  values of the polarization   percentage P\%   {  $\approx$ 0.05  \%\ (\hb) -- 0.12 \%\ (\ha)} lower than the one of the average { adjacent} continuum polarization { ($\approx 1.07$ \%)}. 

}

The profiles of the polarization position angle are shown on     panel 5 of  Fig.  \ref{fig:stokes}.  { In both the \hb\ and \ha\ lines, one can detect a double "swing": the  profiles change within around $\pm$10$^\circ$ relatively to the mean level that the continuum polarization shows, with minimum PA around $-7000$ \kms. These features point  out the presence of the equatorial scattering in the nucleus region, according to \citet{Smith2005}, and the detection of such profiles in both lines is a strong evidence in favor of this model. This inference is consistent with the Pop. B nature of Fairall 9 i.e., with Fairall 9 belonging to the "disk dominated" type-1 AGN \citep{richardsetal02}. Following the paper by \citet{Afa2015} we examine below  these profiles in details to estimate the viewing angle and the black hole mass, and to finally resolve the velocity field of the gas emitting the reflected radiation. }

\section{\label{sec5}Discussion}
    \label{disc}



\begin{figure}
\includegraphics[scale=0.4]{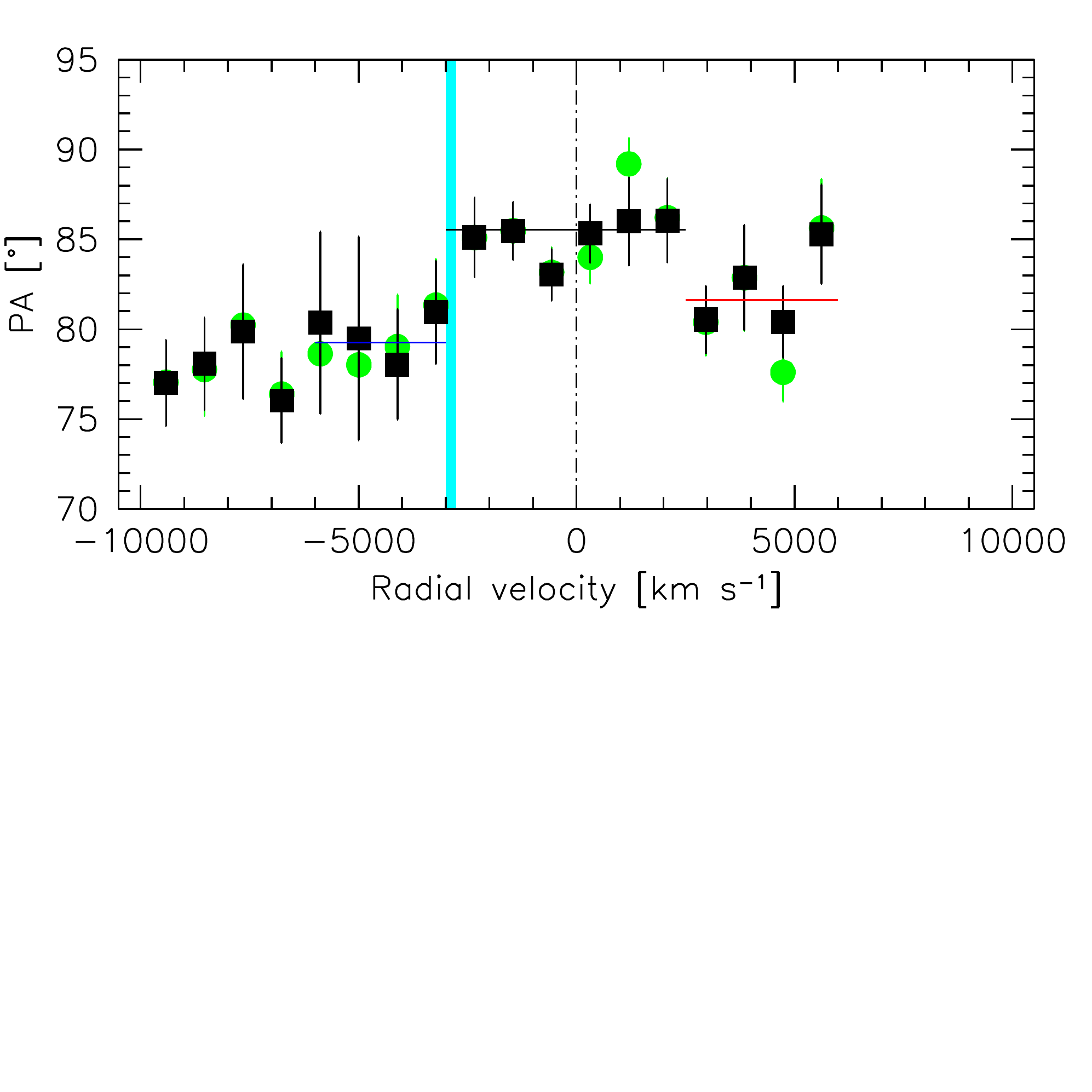}
\vspace{-3.5cm}
 \caption{ The PA as a function of radial velocity, for the average (black squares) and weighted average (green squares) of the \hb\ and \ha\ PAs. Horizontal lines trace the average values of the PA over three ranges in radial velocity. The cyan stripe identifies the range over which a significant change in PA occurs, around $- 2750$ \kms. }
\label{fig:hbha} 
\end{figure}

\begin{figure*}
\includegraphics[scale=0.27]{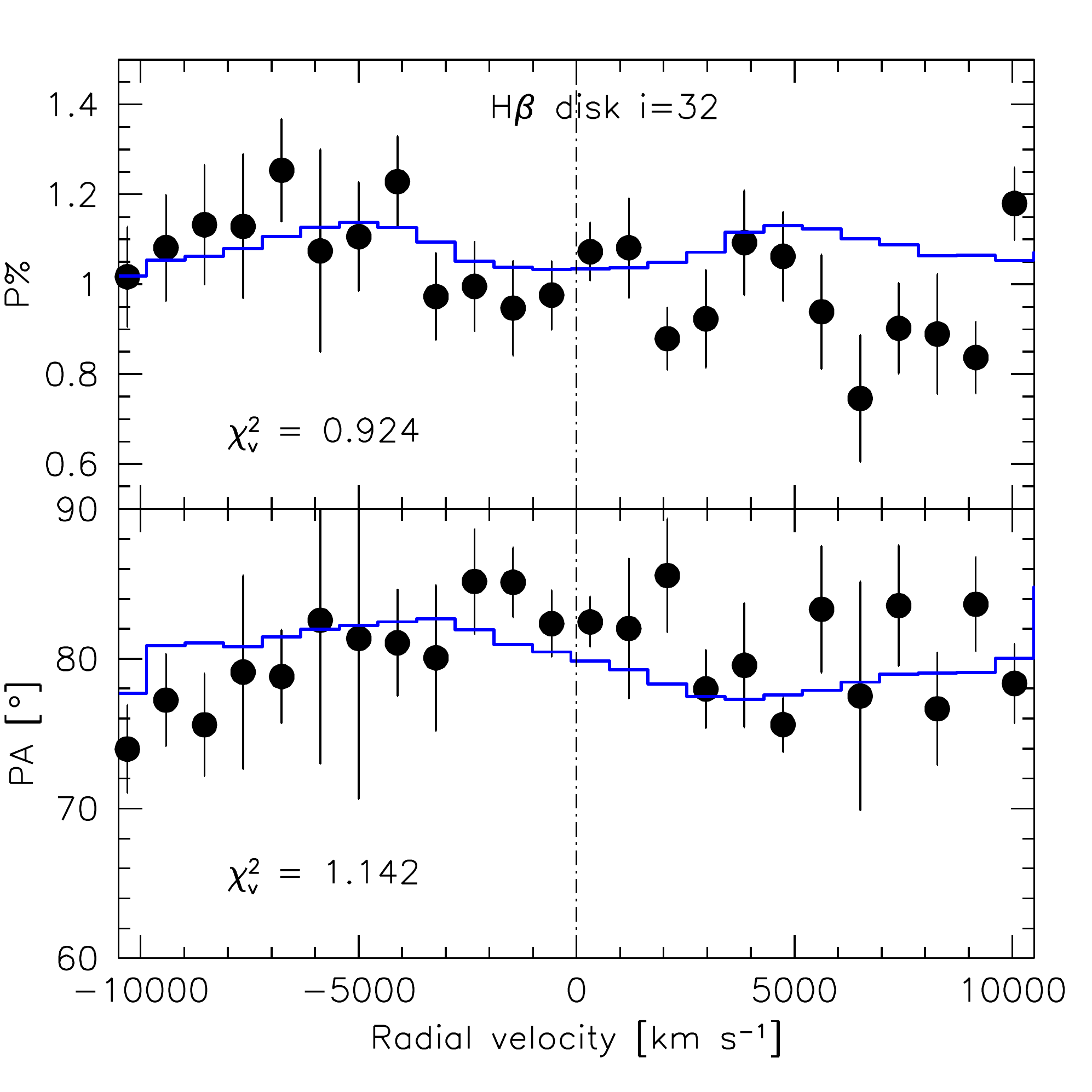}
\includegraphics[scale=0.27]{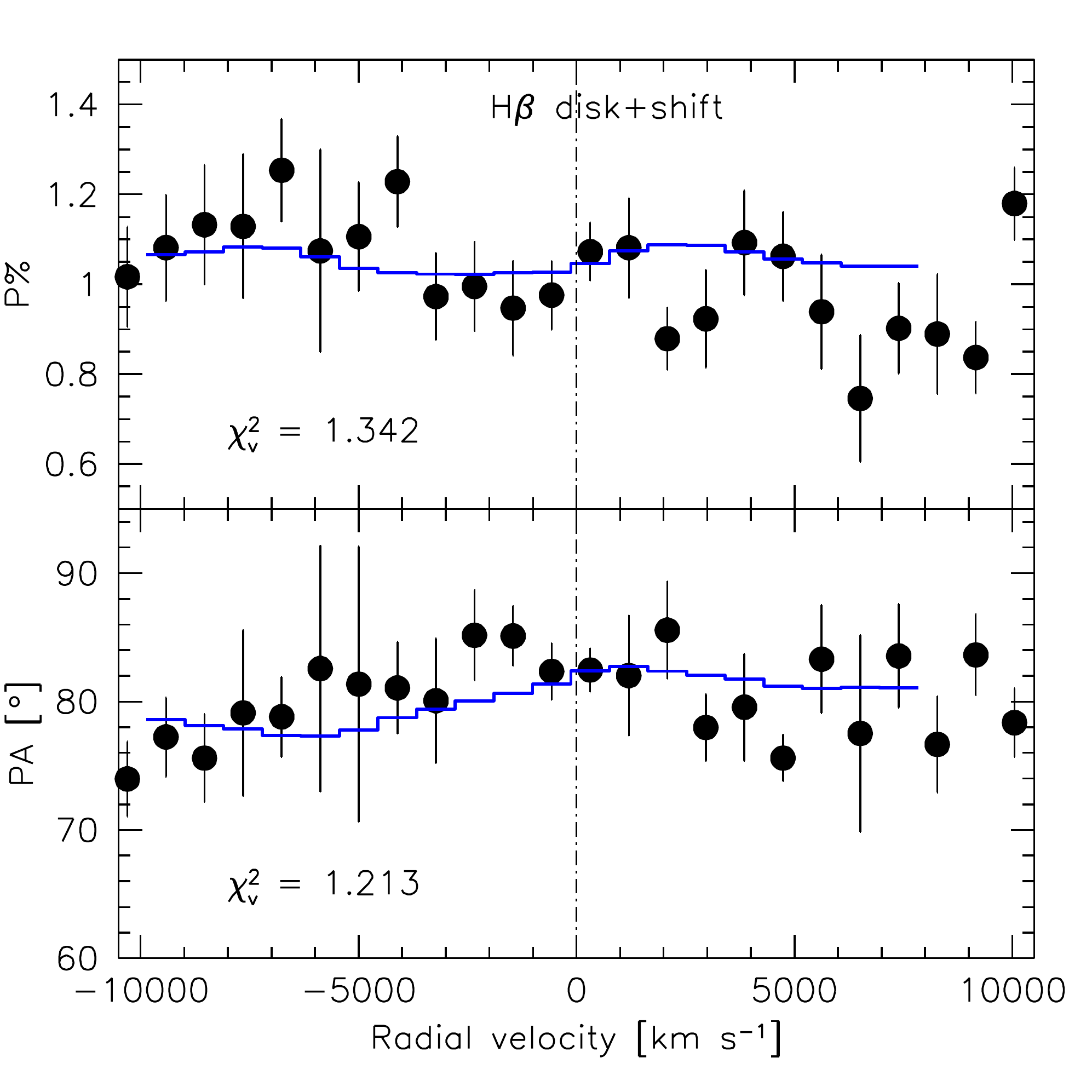}
\includegraphics[scale=0.27]{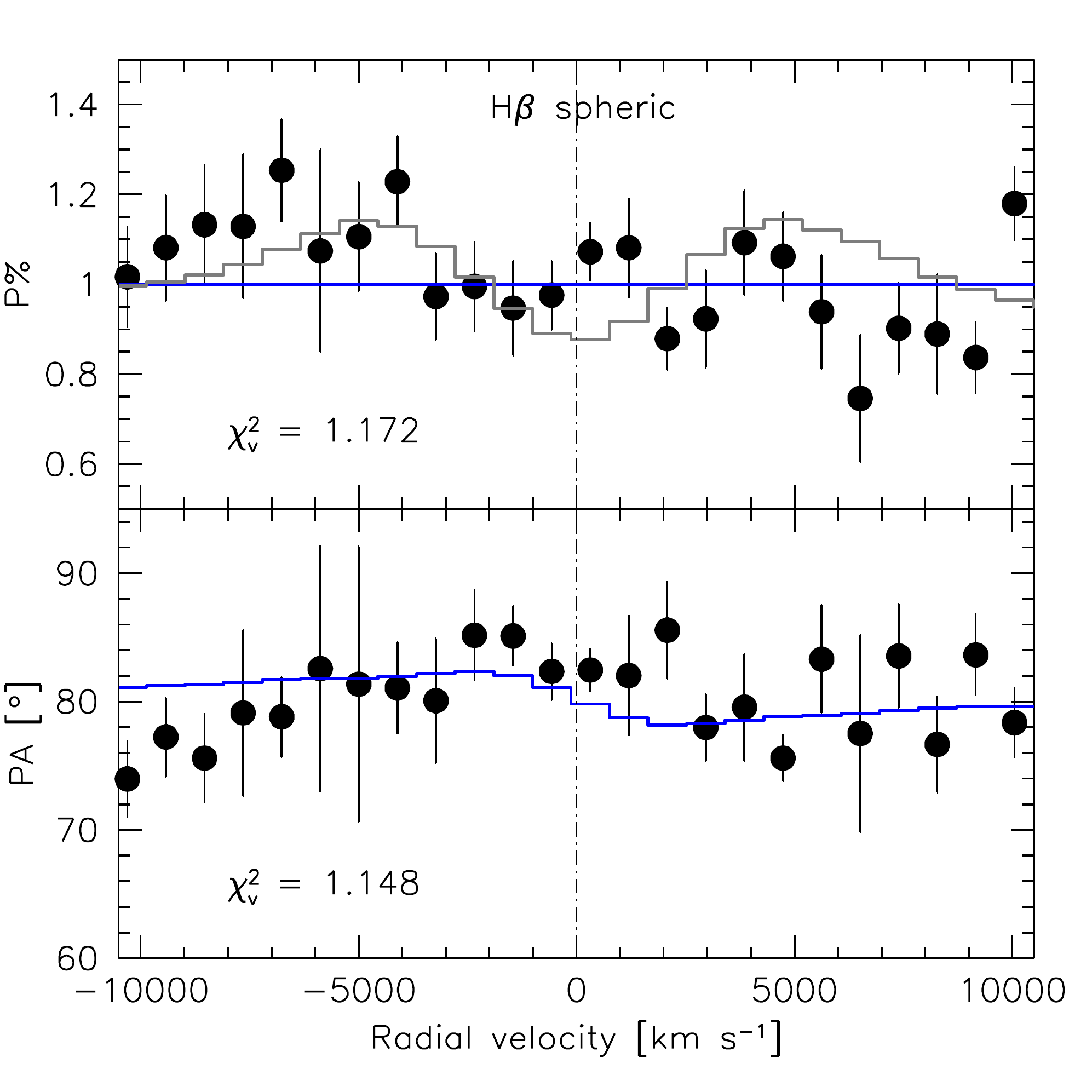}\\
\includegraphics[scale=0.27]{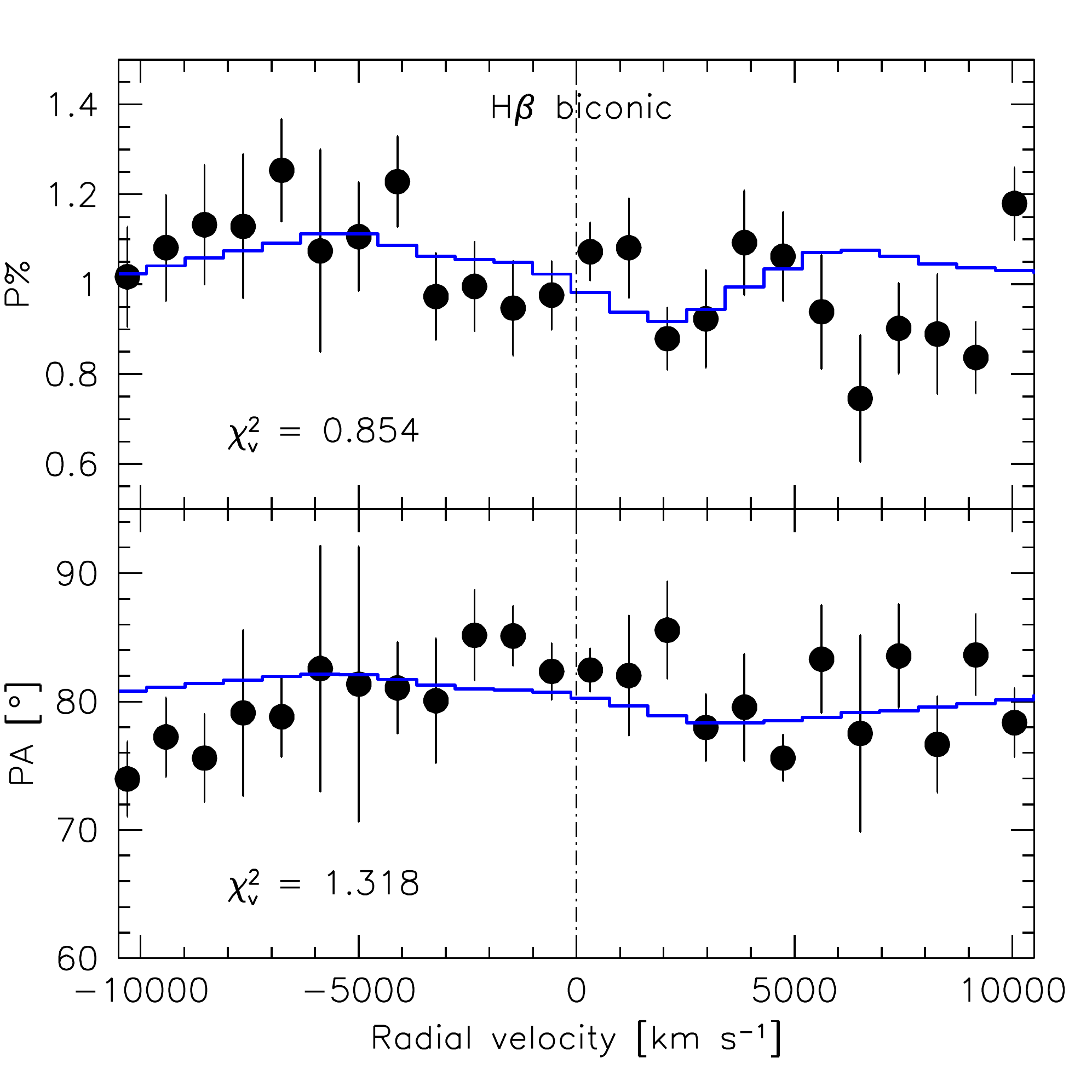}
\includegraphics[scale=0.27]{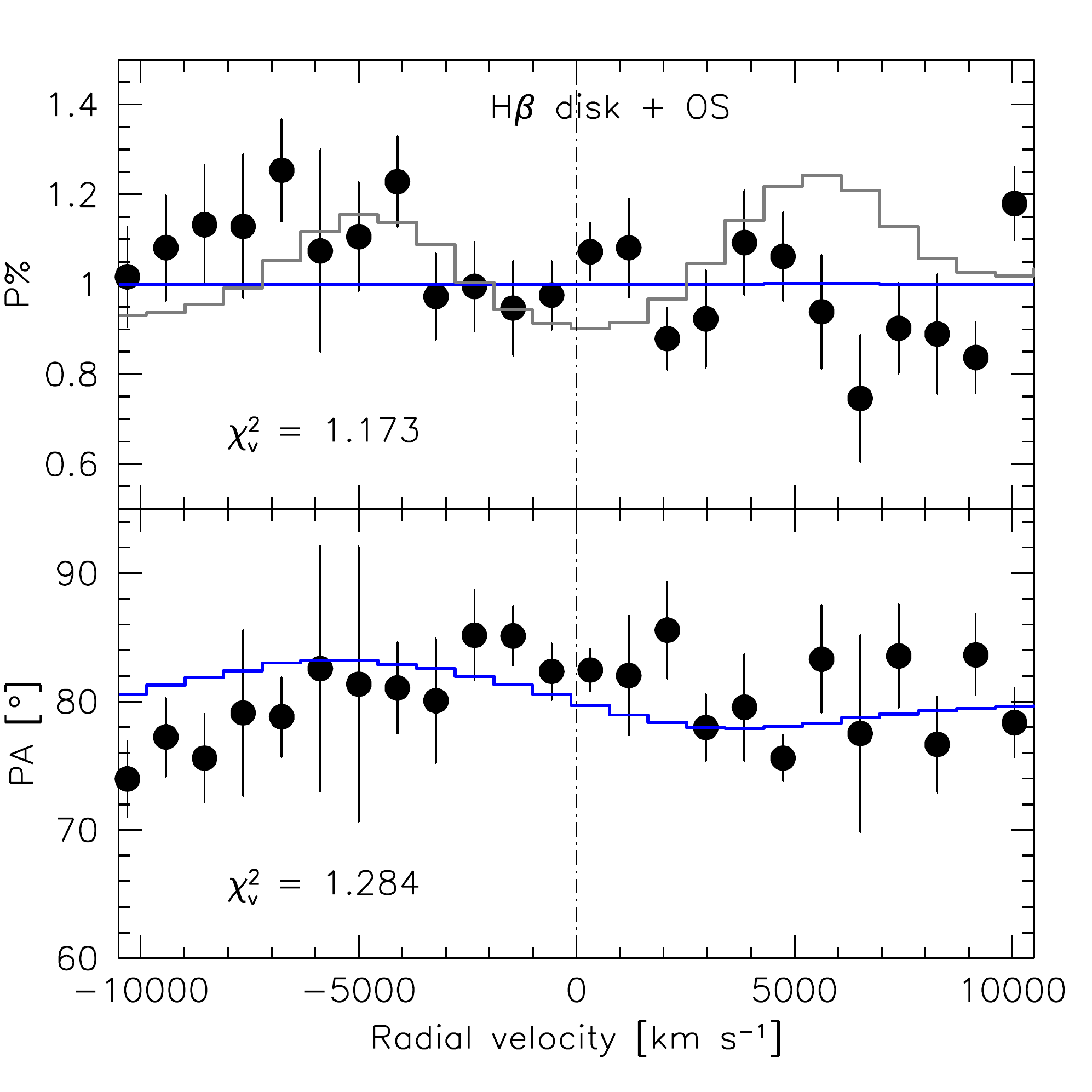}\\
\includegraphics[scale=0.27]{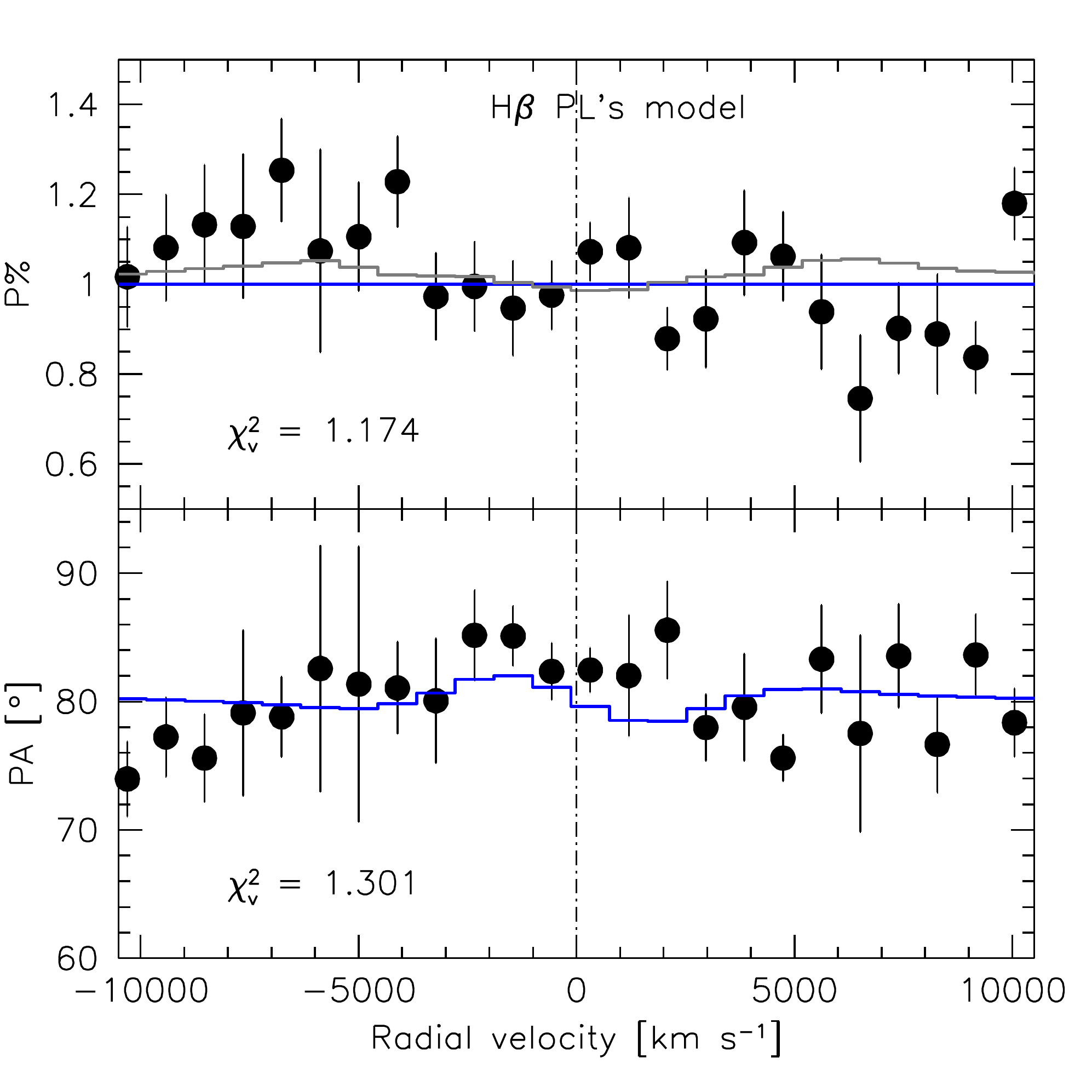}
\includegraphics[scale=0.27]{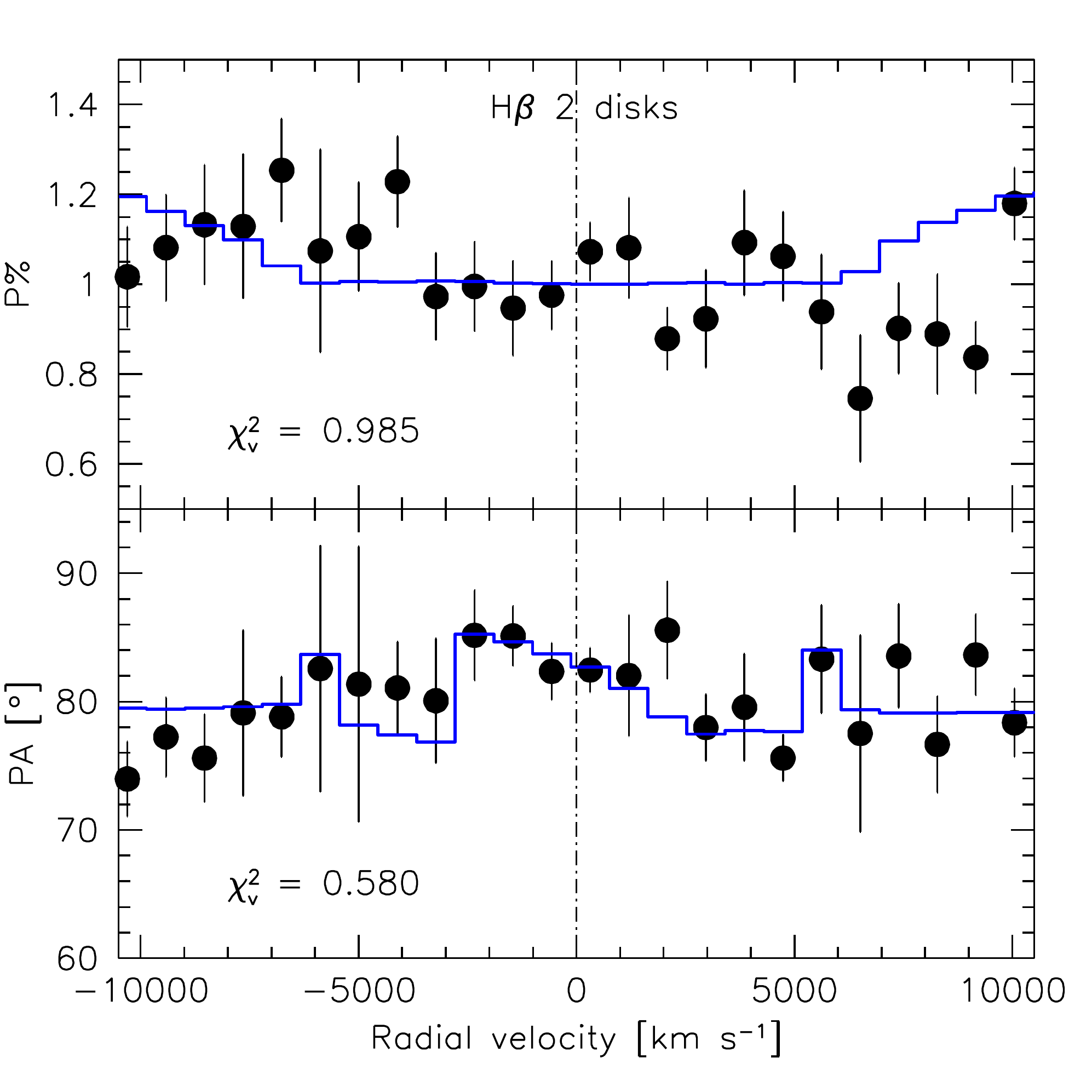}

 \caption{{Comparison between several  ray-tracing models and the observational data for \hb. Each panel is split in half; the top one shows the polarization percentage and the bottom one the polarization PA. Top row, from left to right: results for a single disk geometry and equatorial scattering, for a single disk geometry to which an arbitrary shift has been added, and for a spherical geometry. Second row from top: biconical outflow model, disk with outflow motion of the scatterer by 500 \kms, and a disk + 4000 \kms\ scatterer outflow, with the scatterer overlying the emission disk  as  developed by \citet{liraetal20}. The third row are the results for a two-disk model with the two disks inclined by 30 degrees with respect to each other, with the outer disk observed at 15 degrees. }\label{fig:twodisks}}
\end{figure*}

Appendix  \ref{models} provides \textsc{stokes} and \textsc{skirt} models for several geometric and kinematic configurations. The sketch of Fig.   \ref{fig:model_geometry} shows a Keplerian disk, a polar outflow region, a spherical BLR, and a double disk surrounded by the equatorial scatterer -- a dusty torus.  In the first part of the discussion we will describe the comparison between these and other competing models and the data, beginning with the possibility of polar scattering,  while in the second part we will focus on the analysis of the models that are suggested by the data. For the  quantitative $\chi^2$\ comparison we consider  only \hb\ { up to $+5000$ \kms\ (beyond this radial velocity, the [O{\sc iii}] doublet   destroys the polarization profile of broad \hb). } The \ha\ profile is more heavily affected by the \ha\ NC, by [N {\sc ii}] emission, and by the B band   within $|\delta v_\mathrm{r}| \lesssim 2000$ \kms, right where S/N is higher.  The \ha\ polarization profiles gave always worse $\chi_\nu^2$\ than \hb, but  the main features of \ha\ profile  are qualitatively consistent with the main features of \hb. \ha\ is therefore used for confirmatory purposes.     
The most robust, salient feature that we need to explain is the almost constant polarization degree across the \hb\ and \ha\ profile, which in turns lead to a single-peaked polarized flux profile. The second main feature is the PA profile, whose 0 point is apparently displaced from rest frame. { In this respect, Fig.  \ref{fig:hbha} shows average and weighted average of the \hb\ and \ha\ PA data. A significant change in PA is not occurring  at $v_\mathrm{r} \approx 0$ \kms, but at   $v_\mathrm{r} \approx -2750$ \kms: a Welch $t$-test yields a significance  $ \sim (1 - 3 \cdot 10^{-4})$\ for the change in  PA between the  radial velocity averages over the range  $-$6000 -- $-$3000 \kms and $-$3000 -- $+$2500 \kms\ (blue and black lines in Fig.  \ref{fig:hbha}). The absence of a ``swing" centered at  $v_\mathrm{r} \approx 0$ \kms\ is also visible in Figs.  \ref{fig:stokes} and Fig.  \ref{fig:PA} for both \hb\ and \ha, and their average is reinforcing the result at a statistical level. }

\subsection{\label{sc-region} Polar scattering}

In the previous discussion we have assumed that the scattering region is equatorial, with relatively small inclination with respect to the symmetry plane offered by the accretion disk. However, this assumption may not be fully adequate.  For Pop. B sources like Fairall 9, the inclination is relatively large. It is reasonable to consider the effect of polar scattering as done by  \citet{smith2004}. In the interpretation of \citet{smith2004} of sources showing evidence of both polar and equatorial scattering, the polar scattering region is subdivided into (1) a scattering cone  aligned with the emission disk axis,  but with a very large opening angle as shown in Fig. 10 of  \citet{smith2004};  (2) a spherical wedge at large inclination (but still low enough for the line of sight not to intercept the torus). If the main component of the polar scatterer is distributed in this configuration, we can expect that the width of the polarized  flux feature and of the swing PA pattern will be narrowed, polarization degree lowered, and that PA might be different in the continuum and in the line. 

A combination of polar and equatorial scattering appears,  { in principle} a possibility. This would not have been  a first occurrence  by far.  A combination of polar and equatorial scattering is expected as a general feature of type-1, where the polar scattering electrons are provided by gas in correspondence of the narrow-line region \citep{smith2004}. This could make it possible  to form a central peak in the polarization profile. The inclination estimates  imply that the line of sight may pass close to the edge of the torus. However, Fairall 9 is a very low polarization object, with no sign of significant reddening and highly polarized continuum, suggesting that it should be seen well above the torus (and with the line of sight not passing through the wedge). The source is   different from Fairall 51, the prototypical ``Seyfert on the edge.'' In addition, if a polar scatterer is assumed as in region (1) of \citet{smith2004}, we should see evidence of narrowing in the polarized  flux feature and in the swing PA pattern, which we don't: the swing PA\ pattern is wide.

\subsection{Model inter-comparison}

 { Fig.  \ref{fig:twodisks} shows the data superimposed to the best fits from the models discussed below. The minimum normalized $\chi_\nu^2$ values obtained from the model best fits to the data are reported in each panel. The continuum polarization has been added and the polarization \%\ predicted by the model has been scaled by a free factor to obtain a minimum $\chi^2$. They are computed in the range $-10000 \le v_\mathrm{r} \le 6000$\ \kms. Beyond 6000 \kms, there is evidence of depolarization associated with the narrow [O {\sc iii}] emission.  Considering that the number of degrees of freedom is $16$, for most models, an F-test based on the normalized $\chi_\nu^2$\ ratios for the different models would require a minimum $F(1,2) = \chi_{\nu,1}^2/\chi_{\nu,2}^2 \approx 1.268$\ for detecting a difference that is significant at a 1$\sigma$ confidence level. The minimum $\chi_\nu^2$ is often consistent with a flat or almost flat behavior, which might imply a very low polarization degree associated with a low optical depth of the scatterer.  To show the behavior expected for each model, the grey histogram line in Fig.  \ref{fig:twodisks}  traces the model predictions in an arbitrary scale.}

\subsubsection{Single disk model + equatorial scatterer}

Having excluded polar scattering, we turn to the Keplerian disk and equatorial scatterer model that we implicitly applied for the \mbh\ estimate. This is the model that has been successfully applied to several tens of type-1 AGN \citep{Afa2019}. It  shows a  typical double-swing feature of PA with the $\mathrm{PA}_\mathrm{max}$ amplitude decreasing as the viewing inclination increases.   The polarized line is broader than the line in natural light and the degree of polarization shows double-peaked profiles with maxima in the wings and minimum in the core (Fig.  \ref{fig:Fairall9_STO_SED.pdf}). However, the polarized  flux profile of Fairall 9 is single peaked for both \hb\ and \ha.\ The polarization percentage across the profiles changes little, and   may indicate a different  situation from the simple case of the rotational motion in a flat disk and equatorial scatter.   { Fig.  \ref{fig:twodisks} shows that the agreement between data and model is fair.  The $\chi^2$\ is not very high  for both the polarization \%\ and the polarization PA.  The single disk profile seen at an inclination of $\approx 30$\ degrees is consistent with the data for both the P\%\ and the PA, supporting the presence of rotational motions.

No  improvement is obtained if a blueshift of about $\approx 2500$ \kms\ is imposed to the disk model. This ad hoc modification is suggested by the PA profile that is not symmetric around 0, as already shown in Fig.  \ref{fig:PA}. A shift would significantly lower  the \ha\ PA $\chi^2$. However, for \ha\ the improvement in the polarization angle is not accompanied by a corresponding agreement in the polarization \%\ profile, and for \hb\ there is no significant improvement. } 



\subsubsection{Spherical BLR}

The spherical BLR model involves an isotropic velocity dispersion along with the spherical geometry. The  typical double-swing feature of PA\ is visible also in this case, and the models of the single Keplerian disk and the spherical BLR give similar results.  The agreement is again fair. {A shift to the blue by $\approx - 2000$\kms\ improves the polarization \%\ profile, at the expense of a worsening of the PA $\chi_\nu^2$.}

\subsubsection{Outflows}
\label{nv}

{We considered a model for which the BLR geometry follows a bi-conic outflow with constant velocity\citep[Fig.\, \ref{fig:model_geometry}, top second panel, ][]{Zheng1990,PM96,Corbett2000}. Such model produces a clear double-peaked unpolarized lines, while the polarized line is highly asymmetric with blue part being dominant towards intermediate viewing angles, while the red part is dominant towards pole-on view. The Stokes parameter $U$ is anti-symmetric with respect to the line center, while $Q$ is asymmetric, which results in asymmetric double-swing PA profiles.} 
  The biconic model provides the minimum $\chi_\nu^2$\ among all models, but the fit of the PA profile is poor, and disfavored at a 1$\sigma$ confidence level.  {   Biconical outflows require high Eddington ratio and may therefore be specific of sources accreting at higher rate than Fairall 9. }
 
 { Pop. B sources also show evidence of outflows as discussed in Section  \ref{results}, even if less powerful  than in Pop. A. We consider two  alternatives. The first, outflow in the scattering medium located at the inner edge of the torus (in the previous models the scatterer was assumed stationary) yields a poor   $\chi^2$. As in the case  of the single disk, the model predicts two peaks in the \%\ of polarization that are not observed.  The model with an outflowing scatterer overlying the disk emission could be interpreted as a disk wind, perhaps driven by magneto-hydrodynamical forces \citep{emmeringetal92}. Unlike the cases studied by \citet{liraetal20}, the agreement with our data is not good:  the  $\chi_\nu^2$\ of the P\%\ profile is consistent with unity (even if the minimum  $\chi_\nu^2$\ obtained for a flat profile), but the PA profile is not favored. 
}

\begin{figure}
    \centering
    {\includegraphics[width=0.45\textwidth]{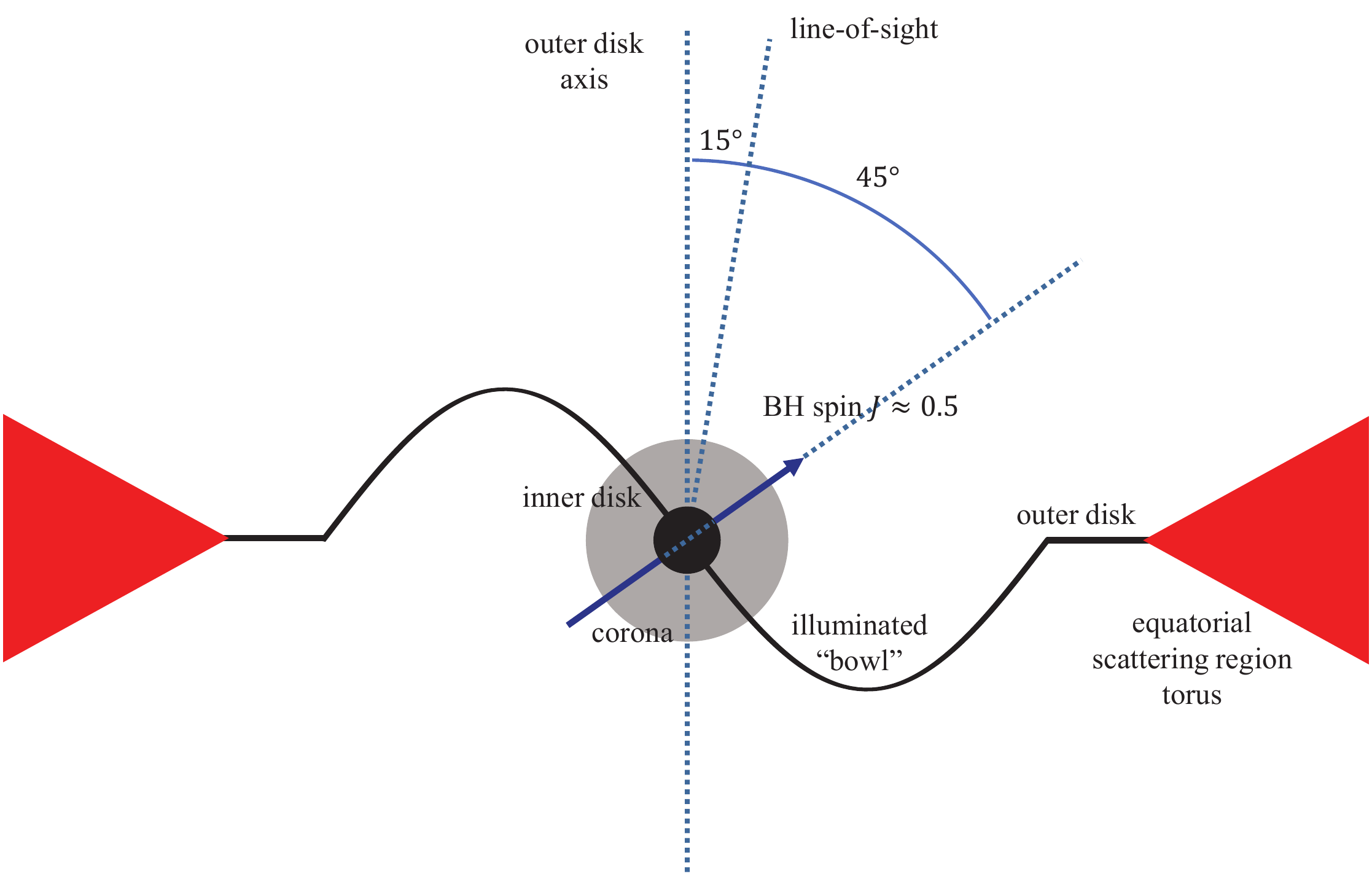}}
\caption{Sketch (not drawn to scale; the
 { curvature of ``bowl'' { is} greatly exaggerated,  and there should be no discontinuity between the outer and inner disk, see Fig.  \ref{beta_J})} illustrating the warped disk geometry suggested in this paper.   \label{fig:sketch}}
\end{figure}

\subsubsection{Double  Keplerian disk}

We simplified the warped disk with a combination of two disk-like BLRs at two inclinations (the two-disk model): the inner disk corresponding to region (2) is rotated by an angle of $30^\circ$ around the $x$-axis with inner and outer radius of 1000 $r_{\rm g}$ to 2000 $r_{\rm g}$. The outer disk (region 1) is situated in the asymptotic  plane that contains the equatorial plane of the torus (Fig.  \ref{fig:sketch}) and extends from 2000 $r_{\rm g}$ to 10000 $r_{\rm g}$.  Region 3 is represented as the point-like source of the continuum. The models of Fig.  \ref{fig:fit} { provide} qualitative confirmation of the feature seen in the Fairall 9 spectrum. Fig.  \ref{fig:twodisks} shows the polarization parameters predicted for two inclined disk configuration (third row from top).  It is interesting to note that the inclination of the inner disk of the model (45$^\circ$) is consistent with the angle  estimated  by \citet{Loh2012} using X-ray spectra($\sim 48^\circ$). The warp geometry might imply that the innermost disk emitting the X-continuum and the inner disk in the model may be facing each other (Fig.  \ref{fig:sketch}).

The main features observed in the polarization profile of Fairall 9 are recognizable in the models. 
\begin{enumerate}
    \item The polarized flux in Fig.  \ref{fig:stokes} shows a centrally peaked structure. The two-disk models shown in the Figs.  \ref{fig:Fairall9_14_SED.pdf} and  \ref{fig:Fairall9_STO_SED.pdf} of Appendix  \ref{models} show the sum of two concentric ``rings'' and clearly represent an oversimplification with respect to the reality of a warped disk. A real warped disk can be assimilated to a sequence of narrow rings with inclination progressively closer to the one of the Kerr black hole equatorial plane, easily producing a single peaked polarization profile. The  point here is that   {   a flat polarization percentage profile within $\pm 5000$ \kms\  yields a polarized flux profile that is centrally peaked as the profile in natural light (panels 5 of Fig.  \ref{fig:stokes}). This behavior is qualitatively consistent  with the polarized flux profile expected for a warped disk, and}   is not consistent with the one expected from a { single} rotating disk and an equatorial stationary scatterer.  
    \item  Negative $\Delta \mathrm{PA}$\ in the radial velocity range from -10000 to -3000 \kms\ are predicted by models and are qualitatively consistent with the change observed in the data (bottom panel of Fig.  \ref{fig:stokes}). 
    \item The shape of the PA around 0 \kms\ is predicted by the model if the disk at $\phi \approx 0^{\circ} - 45^{\circ}$, Fig.  \ref{fig:twodisks}  can also explain the observed PA shape that suggests a change in the sign in the polarization angle  at non-zero radial velocities.   
\end{enumerate}

 { Fig.  \ref{fig:twodisks} shows the model overlaid to the data point. The $\chi_\nu^2$\ is the lowest among the different model, and  best agreement is obtained for the  azimuthal angle $\phi =15^\circ$  case. }{A  $\chi_\nu^2 \approx 0.99$\ is obtained for the P\%\ of the 2 disk model. It is just a factor 1.15 larger than the  $\chi_\nu^2$\ obtained for the biconic model and, according to the $F$-test criterion, not significantly different.  The PA profile of the two-disk model  has a significantly lower  $\chi_\nu^2$\ with respect to all other cases, at least by a factor $\gtrsim 2$, implying that the difference is significant at a confidence level of more than $1\sigma$, and close to 90\%.    }

The innermost part of the BLR is  expected to be seen at  inclination $\gtrsim  30^{\circ}$\ (and significantly different from the 10 -- 15 degrees inferred for the outer disk), because of the relatively modest $\Delta \mathrm{PA}$ in the swing that implies relatively high values of the viewing angle. A warp can indeed change the viewing angle of the line emitting region in the innermost part of the BLR, lowering the swing amplitude. Fig.  \ref{fig:Fairall9_STO_SED.pdf} shows that  the amplitude of the swing is decreasing with viewing angle \citep{Smith2005}.  { If these effects are taken into account, the \mbh\ derived from spectropolarimetry might be underestimated.} 

However, the polarized flux predictions consistent with a  warped disk and the agreement with the two-disk model  PA profile {\it suggest} that a two-disk system or a warped disk might be the most appropriate models.

\subsection{ { A transient second disk}}

 { The model that produces best agreement with the data involves two disks at different inclinations with respect to the line of sight. A tidal disruption event (TDE) could have given rise to a second accretion disk. Fairall 9 has been   described as a changing look AGN in recent papers. It has passed from a very low state in the early 1980s \citep{kollatschnyfricke85}; however, in the low state the broad lines almost never disappeared. Although TDEs are relatively short lived, and destined to fade in the course of a few years, there are claims of longer events \citep{linetal17}. In the case of Fairall 9, the $V$\ observations of the ASAS-SN \citep{shappeeetal14} indicate remarkable photometric stability over the period May 2014 -- Sept. 2018, with an average $V \approx 13.77$\ mag and a dispersion of just $\approx 0.07$\ mag.  We suggest that the strong change occurred in the 1980s might be a ``long-term event of sustained accretion" \citep{trakhtenbrotetal19}. These events might be typical of AGN, and especially among Pop. B sources accreting at relatively low rates.  }

\subsection{A warped disk}
\label{warp}

 { Apart from a second disk due to a TDE, the only mechanism known to us that may yield a change in disk orientation on the spatial scale of the inner BLR is Lense-Thirring precession, produced by the misalignment between the spin of the black hole and the angular momentum vector of the accreting gas \citep{bardeenpetterson75}.}

The expectations for  a warped structure involve three main regions: (1) an outer, asymptotic disk with  a well-defined tilt angle of the plane of the disk relative to the equatorial plane of the black hole $\beta_0$; (2) an intermediate region where inclination $\beta$\ can be much larger than $\beta_0$; (3) a region within $\lesssim 100$\ gravitational radii where the disk is in the equatorial plane of the rotating black hole.  The third region is probably too hot to account for the observed emission line spectrum. It is more likely to host the X-ray corona \citep{begelmenetal83,haardtmaraschi93,rozanskaczerny00}, and be the source of most UV and X continuum.  See Fig.  \ref{fig:sketch} for a schematic representation of a warped structure tentatively adjusted to the observational constraints on Fairall 9.

{Modeling the exact geometry of a warped disk is difficult since   three Euler angles are needed to define the orientation of a warped structure in space. } In the warped disk scenario, the inflection on the natural line profile of \ha\ roughly separates two regions: the outer one (region 1), Keplerian, still in the plane of the accreting gas, and an innermost one more  inclined, exposed to the full strength of the AGN continuum, and producing the high-ionization VBC (region 2; the illuminated ``bowl'' of Fig.  \ref{fig:sketch}; see \citet{bachev99} for a computation of the detailed illumination and self-shadowing patterns in a warped disk geometry). Assuming \mbh\ $\approx 2 \cdot 10^8$ \msun, the radial BLR distance measured from reverberation mapping corresponds to $\approx 1700 r_\mathrm{g}$. A change of inclination can be induced, on a spatial scale of $\sim 10^3 r_\mathrm{g}$\ by Lense-Thirring precession \citep{bardeenpetterson75}.

\begin{figure}
    \centering
    \includegraphics{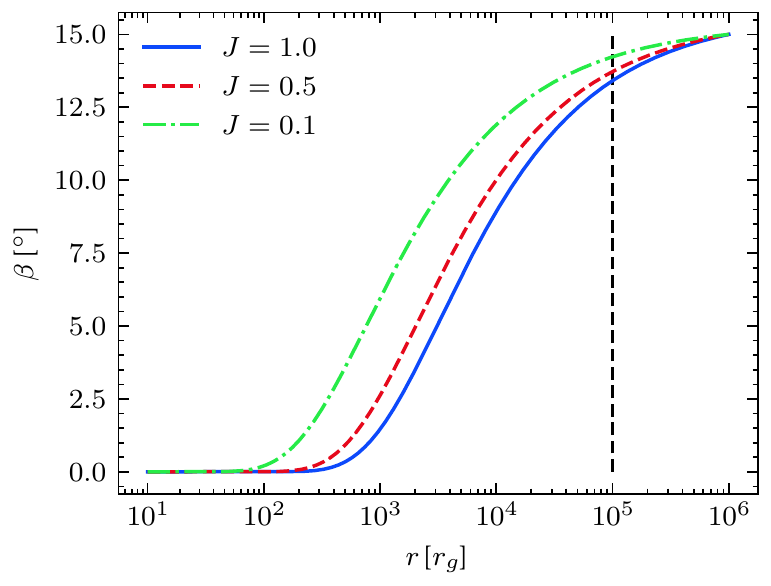}
    \caption{The solution of Eq. (9) in \citet{bardeenpetterson75} with different angular momentum J in units of \mbh$^2$. The tilt angle $\beta$ is normalized to 15$^\circ$ at $10^6r_{\rm g}$. Other parameters in the equation are set to unity.}
    \label{beta_J}
\end{figure}

The trends of Fig.  \ref{beta_J} represent the behavior of the  inclination angle of the warped disk $\beta$ as a function of the radial distance in units of gravitational radii for $J = 0.1, 0.5, 1$, and are  expected for Lense-Thirring precession as computed by  \citet{bardeenpetterson75}. The Lense-Thirring scale is consistent with the linear scale derived from $r_\mathrm{BLR}$. It is also interesting to see that we expect a strong increase in the inclination of the disk at a few hundred gravitational radii from the black hole, and that inclination of the disk start deviating from the one of the equatorial plane of the black hole at smaller distances for lower $J$.  This is in turn expected to increase the direct illumination of the disk by the central UV/X continuum source, known to be very low in the case of a geometrically thin disk \citep[see e.g., Section 5.10 of][]{franketal02}.  

The gravitational + transverse redshift in the VBC goes as $\delta z \approx \frac{3}{2} r_\mathrm{g}/r$\ \citep{bonetal15}. Therefore, the VBC emission could be strongly affected by gravitational redshift: the  rotational velocity scales with $r^{-\frac{1}{2}}$, implying that $r \approx (c/v)^2 r_\mathrm{g}$. For a rotational velocity $v \sim 15000$ \kms, $r \approx 450 r_\mathrm{g}$, and we obtain a gravitational + transverse redshift  $\sim 1000$ \kms.   Fig.  \ref{beta_J} shows that the $\beta$\ value remains close to 0 up 400 $r_\mathrm{g}$\ for the case of a maximally rotating black hole  \citep[$J \approx 0.994$,][]{thorne74}. If $J < 1$, as mentioned, the maximum radius at which the disk plane lies in the equatorial plane of the black hole should be lower. Therefore, gas at very small radii could be more efficiently illuminated, and the emitted radiation should be even more strongly affected by gravitational redshift  with $J<1$\ than with  $J\approx1$.\ { A Suzaku measurement of the Fairall 9 spin yields $J \approx 0.60 \pm 0.07$ \citep{schmolletal09}.  } 


A Kerr black hole is in principle capable of producing a rotation of the polarization plane via the ``gravitational Faraday effect'' \citep{ishiharaetal88}. The effect is achromatic, but strongly dependent on the distance from the central black hole: the rotation $\delta \mathrm{PA} \approx \frac{5}{4} \pi m J \cos \theta_0 /r^3_\mathrm{min}$, where $m$ is the mass of the black hole in the natural units, $r_\mathrm{min}$ is a photon  impact parameter, and $\theta_0$\ is the angle between the spin of the black hole and the line-of-sight. Assuming $r \sim 100 m$, $J \sim 1$, $\cos\theta_0 \sim 1$, the  $\delta \mathrm{PA} \ll 1$ degree. Even if recent numerical simulations suggest a larger effect \citep{chenetal15}, a significant $\delta \mathrm{PA}$\ ($\sim$ 10 degrees) is possible only at $r \lesssim  10 r_\mathrm{g}$, i.e., for photons in the X-ray domain.

Invoking a warped disk presents several drawbacks. Warps are expected to be rare, as they require special conditions: rejuvenation of a black hole from gas  whose angular momentum is not aligned with black hole spin.  Large systematic redshifts and blueshifts should be equally possible \citep{wangli12}, as the illuminated bowl of a warped disk would lose axial symmetries \citep{bachev99}.  The large shifts and irregular profiles predicted by kinematical models are relatively rare (but large blueshifts are found; see several examples in \citet{zamfiretal10}, while the Balmer line profiles of Fairall 9 are fairly typical for a Pop. B object). However, the redward asymmetry might be {\em always} associated with gravitational redshift \citep[see in this respect][]{pop95,corbin95}, and a warped geometry may contribute to make it more evident, as the bowl shape should make it possible to efficiently expose the line emitting gas at a few hundreds gravitational radii. 

\subsubsection{Fairall 9: analogies and differences with E 1821+643 and Mark 668}
\label{analog}


{ The broad emission Balmer lines of E 1821+643 show an unusual shape with a highly red asymmetric profile and a large broad line peak  redshifted ($\sim$ 1000 \kms) relative to the narrow lines \citep[see][]{sh16}.  However, E 1821+643 is a striking example of profile reversal in polarized light, where to a prominent redward asymmetry corresponds a prominent blueshifted excess in polarized light  \citep{smithetal02} explained  by the emission of one active component of a binary super-massive BH or a recoiling black hole after  collision} \citep{robinson2010}. As done by \citet{robinson2010}, we identify two components in the BLR, one with 0 or a smaller redshift, and one broader with a larger redshift ($\sim 2000$ \kms). However, considering that the features can be identified in most Pop. B sources (about 50 \%\ of quasars), it seems unlikely that they could be due to a bulk motion of the BLR with respect to the host galaxy. Since the broader feature is ascribed to the inner most part of the emitting regions, profile reversal in natural light (i.e., the redward asymmetry turning into a blue one or profile turning from asymmetric to symmetric or vice-versa) should be expected on timescales of a few years.
The first published spectrum of sufficient quality was obtained in 1977 \citep{hawleyphillip78}; the \hb\ profile is remarkably similar to the one of the observations in 2019. Over a 42 yr time lapse the total displacement is significantly larger than the BLR distance from the central continuum source: $\approx 2.65 \cdot 10^{17}$\ cm, assuming a constant velocity of 2000 \kms. If the original BLR radius was $\approx 1.44 \cdot 10^{17}$ cm,  the recoiling displacement  added to the original radius implies that the ionization parameter  of the gas ``left behind'' in the BLR should be lower by a factor $0.1$, and that there should have been a decrease in the \ha\ and \hb\ BC intensity with respect to the VBC by a factor $\approx 0.38$, assuming the same, typical conditions for the BLR. There is no positive evidence of this decrease (for an account of the variability pattern, see \citet{lubderuiter92}). \citet{kollatschnyfricke85}  reported the almost disappearance of the broad \hb\ emission   between 1981 and 1983, but afterwards the AGN bounced back to its pre-1981 spectrum. Further monitoring on a timescale of a decade should provide a stringent test on the recoiling black hole hypothesis.

 { Mrk 668 (OQ 208) is known since long because of a  redshifted peak by $\sim 2600$ \kms\  in the natural light profile of \hb\ \citep[][and references therein]{gezarietal07}. The peak radial velocity has remained approximately constant since the first report on  the shifted feature, over 40 years ago \citep{osterbrockcohen79}. In this case, the presence of a black hole binary is not supported by the data \citep{doanetal20}.  The spectropolarimetric data are consistent with $b = -0.5$ suggesting  a predominance of Keplerian motion. 
The  $\Delta \mathrm{PA}$ changes sign at $\delta v_\mathrm{r} \thicksim 0$, as in the case of Fairall 9 \citep{Afa2019}. 

A warped disk geometry is affected by self shadowing, as the disk illumination is strongly dependent on the azimuthal angle $\phi$. Single peaks highly displaced to both the red and blue are predicted for the Balmer line profiles in natural light \citep{bachev99,wuetal08}. In the case of Mrk 668, optical Fe {\sc ii} emission shows a consistent shift with the Balmer lines \citep{bonetal18}, providing additional evidence supporting that we are seeing low-ionization emission lines from a virialized, Keplerian system, with the Fe {\sc ii} emission occurring predominately at the outer edge of the disk, as expected \citep[][and references therein]{pandaetal20}.  On the ground of the stability of the profile over a timescale longer than the dynamical timescale of the BLR, and of its spectropolarimetric properties, we suggest that OQ 208 could be considered as a  warped disk candidate.   }

\subsection{\mbh\ and \edd\ estimate for Fairall 9} 
\label{sect-mass}
\subsubsection{\mbh }
{ The details of the \mbh\ computations following the spectropolarimetric and other methods are reported in Appendix  \ref{mass}. 
The various estimates reported in Table  \ref{table2} disagree by a factor almost $ 3$,  within the range \mbh $\approx (1.2 - 3.3) \cdot 10^8 $ \msun. Each method has some strong point but also some difficulties. The spectropolarimetric mass estimate depends on the distance of the scatterer, which in turns depends on scaling laws connecting it to luminosity (Tab.  \ref{table2}). The weighted average  of the  \mbh\ values reported in Table  \ref{table2} from spectropolarimetry is $M_{\rm BH} \approx (1.49 \pm 0.48) \cdot 10^8$ \msun.  
It is doubtful whether the \citet{Afa2019} method can be applied in warped disk geometry. Indeed, the uncertainty in the $a$\ parameter rather large, $\delta a \approx 0.13$\ and $\approx 0.15$\ for \hb\ and \ha, respectively, if compared to the typical uncertainties reported by \citet{Afa2019}, always  $\lesssim 0.12$. However, the spectropolarimetric estimate compares well with the average from the application of the virial relation using $r_\mathrm{BLR}$\ from reverberation mapping\footnote{The RM estimates rely on the virial factor, which is not well constrained in the case of Fairall 9.}, as reported in Table  \ref{table2},  even if it is significantly lower than the estimate obtained from 2 scaling laws based on the \ha\ FWHM, and especially from the width of the Si absorption line at 1.59 $\mu$m, which indicate $M_{\rm BH} \gtrsim 2  \cdot 10^8$ M$_\odot$. 
 We computed a  median and weighted average  black hole mass considering all methods (even if they are not fully independent). { The median $M_{\rm BH} \approx (1.68 \pm 0.48)  \cdot 10^8$ \msun \ (with the semi-interquartile range as uncertainty)} and the weighted average $M_{\rm BH} \approx (1.71 \pm 0.63)  \cdot 10^8$ \msun \ are both  consistent with the \mbh\ derived from spectropolarimetry, and also with the previous determination of \citet{recondogonzalezetal97}.

{ The Pop. B \mbh\ distribution at low-$z$\ covers the black hole mass range $8 \lesssim \log$\mbh $\lesssim 10$\ [\msun], in flux-limited  samples built on long-slit or fiber optical spectroscopic observations \citep{Shenetal2011,PM2003MN}. This result would place Fairall 9 toward the low end of the \mbh\ distribution for Pop. B. However, these samples are subject to a strong selection bias, as sources with a relatively small black hole mass (\mbh $\sim 10^6$\ M$_\odot$)  radiating at low Eddington ratio are increasingly lost even at modest redshift. Very recently, the MANGA survey \citep{yanetal16} has made it possible to detect  low-luminosity, type-1 AGN associated with black hole mass even in the intermediate \mbh\ domain \citep[][Hernandez-Toledo et al. 2021, in preparation]{mezcua.and.dominguezsanchez20}. The loss of Pop. B sources is also relevant at high redshift, as the flux limit of large surveys introduces an Eddington ratio-dependent cut-off at any given \mbh\ \citep{sulenticetal14}.  }}

\subsubsection{\edd\ }

{ The bolometric luminosity of Fairall 9 has been estimated as $L_{\rm Bol} \sim 4.75 \cdot 10^{45}$ erg s$^{-1}$ by integrating the observed SED. The SED for Fairall 9 is provided by \citet{brown+19}, which includes photometric and spectroscopic observations of the AGN ranging from $2.5\cdot10^{-5}$ to $9.6\cdot10^2$ $\rm{\mu}$m. The  Eddington luminosity can be written as $L_\mathrm{Edd} = 1.5\cdot10^{38} ({M_{\rm BH}}/{M_{\odot}})$ erg s$^{-1}$. For \mbh $\approx 2 \cdot 10^8$ \msun, the Eddington ratio is $L_{\rm Bol}/L_\mathrm{Edd} \sim 0.16 $, consistent with Pop. B but close to the limit. The observations of \citet{brown+19} refer to a period when the source was in a high state; assuming the flux of \citet{bentz2009} (also consistent with the average of the \citealt{santos-lleoetal97} monitoring campaign) corrected for the host galaxy contribution would imply a bolometric luminosity $L_{\rm Bol} \sim 1.65 \cdot 10^{45}$ erg s$^{-1}$, and an Eddington ratio $\approx 0.06$. The variability notwithstanding, Fairall 9 has remained within the limit of Eddington ratio associated with  Pop. B  ($\approx 0.2$).}

\section{Conclusion}


Using VLT/FORS2 we obtained  spectropolarimetric observations of the radio-quiet Pop.B source Fairall 9. The measured polarization properties have been used for  inferring constraints on the kinematics and geometry of the BLR as well as on the scattering region.  Our  results on the spectropolarimetric properties of Fairall 9 can be summarized as follows. 

\begin{enumerate}
    \item{ Fairall 9 shows low degree of polarization in correspondence of the Balmer \hb\ and \ha\ emission lines. The  polarized flux profiles of both lines are centrally peaked.}
     
    \item The polarization angle in the center of \hb\ and \ha\ is atypical, in the sense that  the "swing" shape is not as regular as the one seen in most cases by \citet{Afa2019}, and the dynamical center may be shifted { with respect to rest frame}.

 
 \item We consider several scenarios  from the Monte Carlo, ray-tracing code {\tt STOKES} and {\tt SKIRT}: disk-like BLR, bipolar outflowing BLR, spherical BLR, and two-disk BLR.  { The minimum $\chi^2$ in $\Delta \mathrm{PA}$\ is obtained for the two-disk case. }
 
 
 \item The agreement with the two-disk hypothesis indicates that the inner disk may be tilted with respect to the outer disk, and that emission from the VBC may be associated with the inner disk.
 
 \item  { The possibility of a second disk originating from a tidal-disruption event  is not favored on the ground of the photometric { behavior} of Fairall 9 right in the four years preceding the spectropolarimetric observations. }
 
 \item  {  Excluding the possibility of a TDE-produced disk, the most likely hypothesis appears to be  a warped structure driven by Lense-Thirring precession.} Models  support the possibility that a double disk structure may be at the origin of the PA, polarized flux, and polarization percentage behavior. 

 \item Using the width of the polarized broad \hb\ and \ha\ profile, we estimated the inclination and the central black hole mass. The inclination effect plays a significant role in the emission line broadening especially in Pop. B sources. For Fairall 9, the inclination angle cannot be determined as easily as for other sources, but is constrained { around $50^{\circ}$}. The corresponding virial factor is  { $ \approx 2$}. The derived values are consistent with the expected role of the viewing angle along the FWHM axis in the optical plane of the E1 MS \citep{PM2001,pandaetal19}. However, if a warped structure is present, then to consider a single value of the viewing angle could be misleading.  
 
 \item The central black hole mass obtained by using the spectropolarimetric method of \citet{Afa2014} is somewhat lower with respect to the mass estimates from most other methods, which give an average \mbh\ $\sim 2\times10^8$\msun.  {However,  The slope $b$\ is consistent with the value expected for predominance of a Keplerian velocity field.}
 
 \end{enumerate}
 
 If our inferences are correct, the case of Fairall 9 confirms the dominance of Keplerian motions in the BLR of Pop. B sources. The issue of orientation is complicated by the warped structure revealed in this source. The results on Fairall 9 further strengthen the hypothesis that gravitational redshift might be the governing factor of the VBC redshift, due to the easy illumination of the disk gas in a warped geometry 
  { (in this respect, \citet{punslyetal20} suggest that at very low \edd, the effect of gravitational redshift is strong because only the innermost part of the disk is illuminated by the AGN continuum)}. Deep, high-resolution observations of other Pop. B sources  are however needed to test whether the spectropolarimetric results obtained for Fairall 9 may be common and general. 

\section*{Data availability}
The data underlying this article will be shared on reasonable request to the corresponding author.
\section*{Acknowledgements}
{ The authors thank the reviewer whose suggestions and comments helped improve the paper. }BWJ would like to express sincere gratitude to the Astrophysical Observatory of Asiago, Italy, where most of the preliminary work in this paper was efficiently  carried out  thanks to the pleasant and quiet working place and accommodation they generously provided. BWJ and JMW acknowledge financial support from the National Natural Science Foundation of China (11833008 and 11991054), from the National Key Research and Development Program of China (2016YFA0400701), from the Key Research Program of Frontier Sciences, Chinese Academy of Sciences (CAS; QYZDJ-SSW-SLH007), and from the CAS Key Research Program (KJZD-EW-M06). VLA and ES  thank the grant of Russian Science Foundation project number 20-12-00030 "Investigation of geometry and kinematics of ionized gas in active galactic nuclei by polarimetry methods", which supported the spectropolarimetric data analysis. {\DJ}S and L\v CP acknowledge the Ministry of Education, Science and Technological Development of Republic of Serbia for support through the contract  \textnumero451-03-68/2020/14/20002. AdO  acknowledges financial support from the Spanish grants MCI PID2019-106027GB-C41 and the State Agency for Research of the Spanish MCIU through the "Center of Excellence Severo Ochoa" award for the IAA (SEV-2017-0709). Under the same award, PM acknowledges the Hypatia of Alexandria visiting grant. This research has made use of the NASA/IPAC Extragalactic Database (NED) which is operated by the Jet Propulsion Laboratory, California Institute of Technology, under contract with the National Aeronautics and Space Administration.

\bibliographystyle{mnras}
\bibliography{ads1}
\vfill\clearpage
\pagebreak
\appendix
\counterwithin{figure}{section}
\section{The estimated black hole mass}
\label{mass}

\subsection{\label{3.2}The inclination effect}
\label{incl}

In optical spectroscopy, we can observe only the line-of-sight velocity distribution of the BLR. Therefore, it is obvious that inclination of the BLR will play a significant role in the line width \citep{Collin2006}. Spectropolarimetry is a powerful tool to eliminate the inclination effect, since the intrinsic Keplerian velocity distribution can be obtained from the polarized profiles. The effect of inclination can be quantified by assuming that the line broadening is due to an isotropic component plus a flattened component, whose velocity field projection along the line of sight is $\propto 1/ \sin i $, where $i$ is the inclination of the system to the line of sight. Thus, the observed velocity distribution is:
\begin{equation}
\label{eq3.2.1}\delta v_{obs}^{2}=\frac{1}{3}\delta v_{{\rm iso}}^{2}+\delta v_{{\rm K}}^{2}\sin^{2}i
\end{equation}
where $\delta v_{{\rm iso}}$ is an isotropic component, $\delta v_{{\rm K}}$ is the Keplerian velocity, and their ratio should be $\kappa = \delta v_{{\rm iso}}/\delta v_{{\rm k}}\gtrapprox0.1$\ \citep{Collin2006,PM2018}. In our work, we use the lower limitation of the ratio, i.e. $\kappa = 0.1$, and the $\delta v_{{\rm obs}}$ is  the FWHM of the observed line profile from single epoch observations, i.e. $\delta v_{{\rm obs}} = {\rm FWHM}_{\rm obs}$. 

Then  the virial factor which relates black hole mass \mbh\ can be written as:
\begin{equation}
\label{eq3.2.2}M_{{\rm BH}}=f_{\rm FWHM}\frac{R_{{\rm BLR}}}{G}{\rm FWHM_{\rm obs}^{2}}
\end{equation}
where $G$ is the gravitational constant, $R_{\rm BLR}$ is the radius of the BLR. \mbh\ is also related to the Keplerian velocity ${\rm FWHM}_{\rm k}$ as:
\begin{equation}
M_{{\rm BH}}=\frac{R_{{\rm BLR}}}{G}{{\rm FWHM}_{\rm k}^{2}}
\end{equation}
Therefore the virial factor can be written as:
\begin{equation}
\label{eq:virial}
f_{\rm FWHM} = \frac{{{\rm FWHM}_{\rm k}^{2}}}{{\rm FWHM_{\rm obs}^{2}}} = \left[\frac{1}{3}\kappa^{2}+\sin^{2}i\right]^{-1}
\end{equation}

In the Fairall 9 case, { the possible concurrence  of a more complex geometry such as disk warping makes the interpretation of the width of the polarization profile uncertain. In order to measure the velocity dispersion associated with the Keplerian motion, we applied the decomposition technique (as mentioned before) to the \hb\ and \ha\ polarized flux (Fig.  \ref{fig:hb_pol}). To estimate  $f$, here we consider two different velocity FWHM of the \hb\ and \ha\ profile in natural light: one associated with BC only and one with BC+VBC.  The corresponding inclinations and virial factors are reported in Table  \ref{tab:inc}. The BC FWHM is the parameter expected to be more strongly affected by orientation, and the $f$\ value derived for \hb\ BC is the one used for the \mbh\ computation.  }



 {

\begin{figure}
    \centering
    \includegraphics{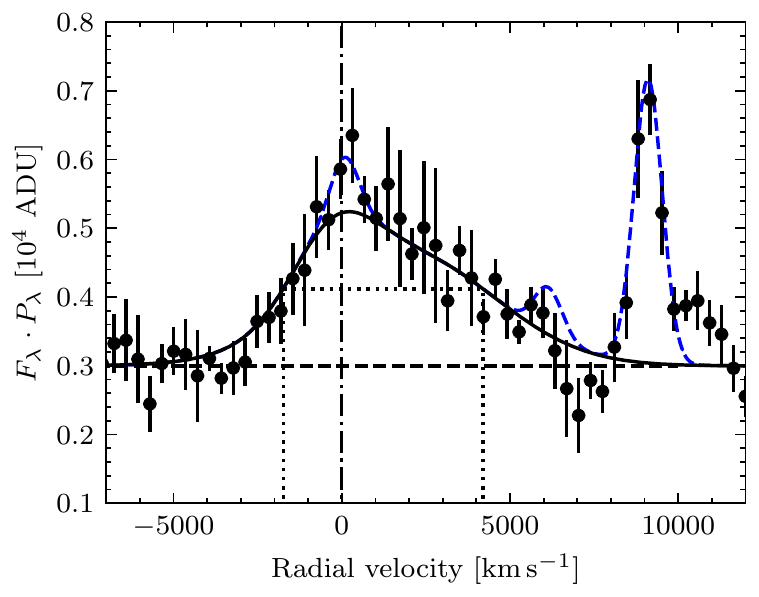}
    \includegraphics{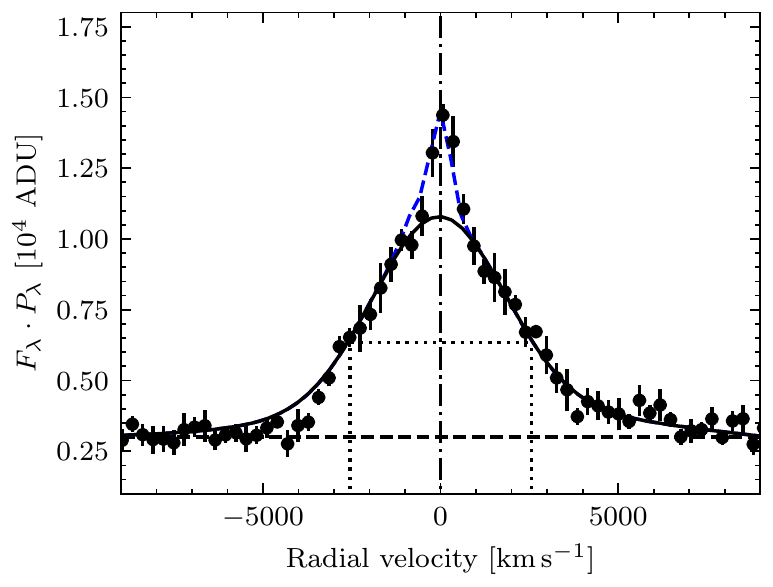}
    \caption{{ Top panel: black dots with error-bars at 1$\sigma$ confidential level represent the polarized flux for \hb. Black solid line shows the best fitting for the total broad polarized profile with two Gaussian components, with dotted lines showing the FWHM. Blue dashed lines represent the narrow polarized components. The continuum level is plotted in black dashed line. The zero radial velocity is marked with dot-dashed line. Bottom panel: same as top panel except for \ha\ polarized profile.}}
    \label{fig:hb_pol}
\end{figure}


\begin{table}
\caption{Inclination and virial factor estimation}
    \centering
    \begin{tabular}{lcccc}
    
    \toprule
      Line component & FWHM($F_\lambda$) & FWHM($P_\lambda\cdot F_\lambda$) & $i$ & $f$  \\
      & (\kms) & (\kms) & ($^\circ$)& \\
      (1) &(2)&(3)&(4)&(5)\\
     \midrule
        \hb\ BC & 4010 & 5925 & 42.4 & 2.18\\
        \hb\ BC+VBC & 5307 & 5925 & 63.3 & 1.25\\
        \ha\ BC & 3848 & 5107 & 48.7 & 1.76 \\
        \ha\ BC+VBC & 5050 & 5107 & 80.8 & 1.02\\
    \bottomrule
    \end{tabular}
    
    \label{tab:inc}
    \begin{minipage}{\linewidth}
    \vspace{0.2cm}
    { \textit{Notes}: Col.  (1), emission line and components in natural light used for the inclination and virial factor estimation. Col.  (2), FHWM of the components in natural light. Col.  (3), FWHM of the total broad profile in polarized light. Col.  (4) and (5), inclination and corresponding virial factor estimated with Eq.  \ref{eq:virial}, respectively.}
    
    \vspace{0.2cm}
    \end{minipage}
\end{table}

}

\defcitealias{peterson2004}{P+04}
\defcitealias{bentz2009}{B+09}
\defcitealias{PM2003}{M+03}
\defcitealias{VP2006}{VP06}
\defcitealias{GreeneHo2005}{GH05}
\defcitealias{Shenetal2011}{S+11}
\defcitealias{Assef2011}{A+11}
\defcitealias{Koshida2014}{K+14}
\defcitealias{Afa2015}{AP15}
\defcitealias{Afa2019}{A+19}
\defcitealias{feigelsonetal92}{F+92}
\defcitealias{Whittle92}{W92}
\defcitealias{Oliva95}{O+95}
\defcitealias{Zuetal2011}{Zu+11}
\defcitealias{McConnell2011}{MC+11}

\subsubsection{Spectropolarimetric method}

\begin{figure*}
\centering
\includegraphics[scale=0.45]{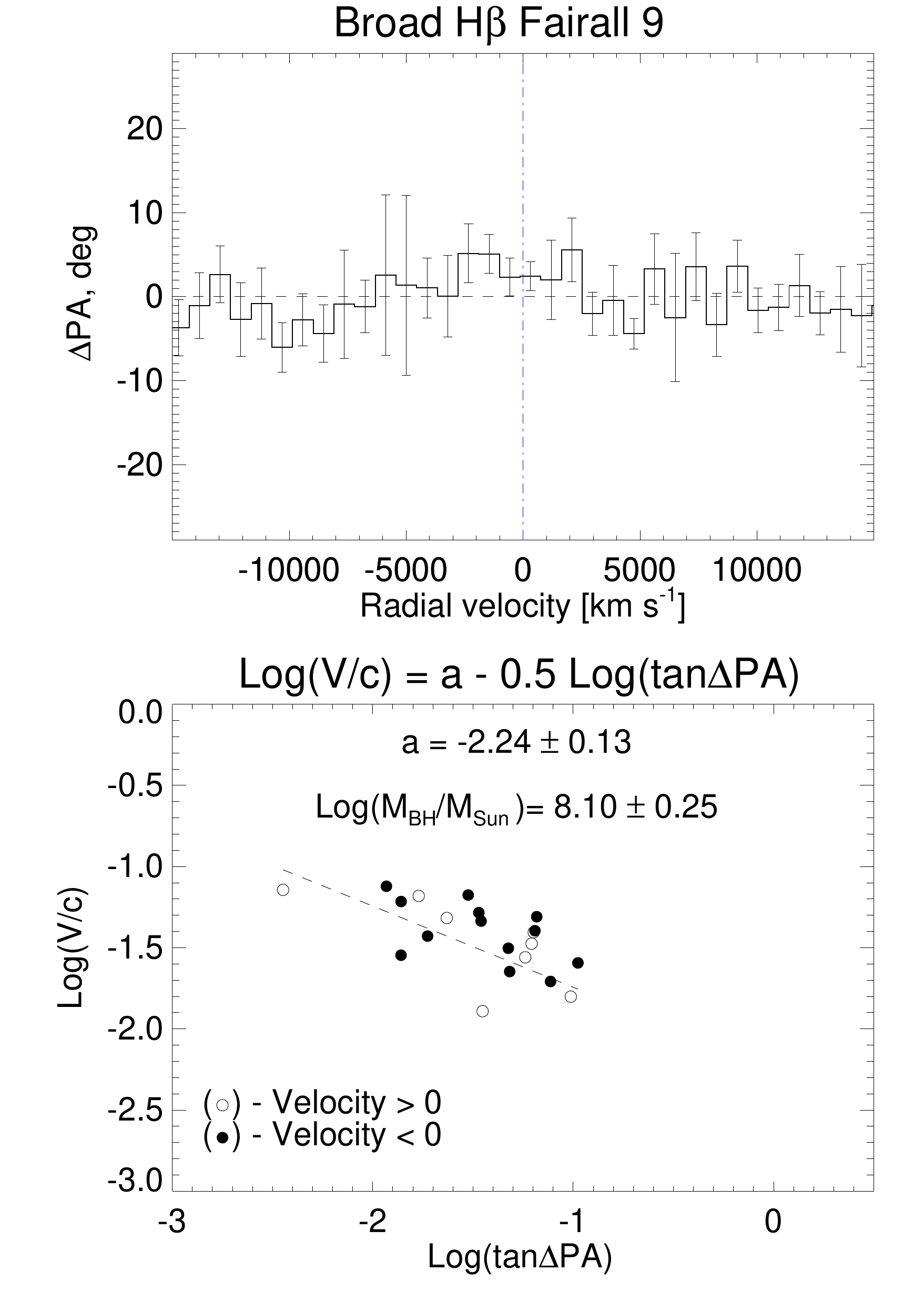}
\includegraphics[scale=0.45]{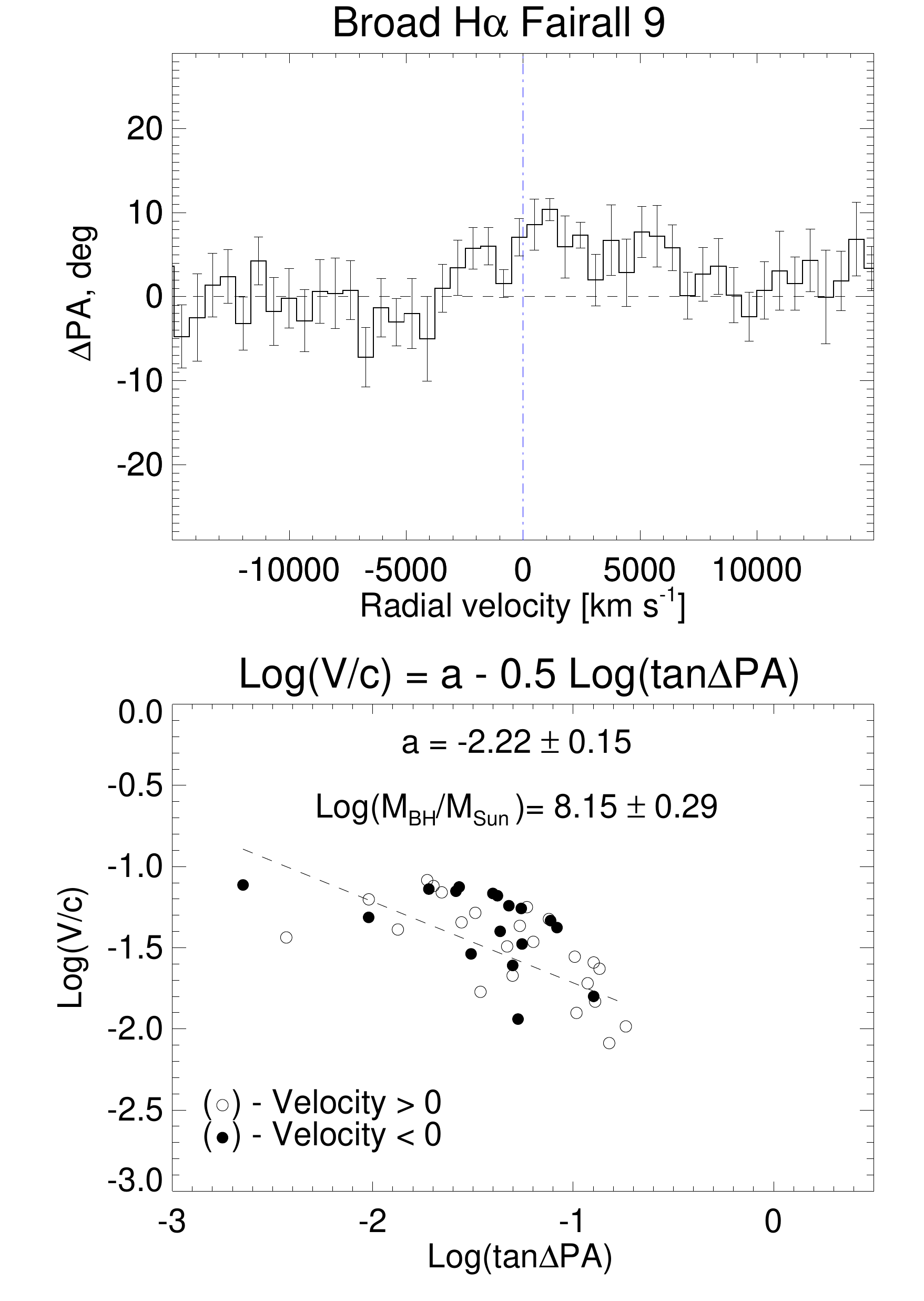}
\caption{\label{fig:PA}  { The polarization angle profiles for \hb\ (left) and \ha\ (right) lines. On the upper panels the deviation of the polarization angle to the mean value (horizontal dashed line) is given. The maximum $\Delta \mathrm{PA}$ is 6$^\circ$ for \hb\ and 10$^\circ$ for \ha. The vertical dashed lines mark the center of the emission lines. The grey shaded area show the contamination of the [O {\sc iii}] emission for \hb\ and the atmospheric B band absorption for \ha\ respectively. The bottom panels show the relation of $\log(v_\mathrm{r}/c)$ v.s. $\log(\tan\Delta \mathrm{PA})$. The \mbh\ shown here is estimated assuming $R_{\rm sc}=215$ light days.}}
\end{figure*}

Several methods are being presently used to estimate the central black hole mass, starting from Eq.  \ref{eq3.2.2}.  { We consider first the} spectropolarimetric data that can be used to estimate the black hole mass from the ``mark'' of Keplerian motion on PA, in a way that is independent from the viewing angle \citep[see][for a detailed description  of the method]{Afa2015}.   In short, a rotating Keplerian BLR will produce a polarized light by equatorial scattering, where the PA of the polarization changes in a way that is correlated to the velocity field of the BLR. The relation between the velocity and the PA is \citep{Afa2015,Afa2019}:


\begin{equation}
\log(\frac{V_{i}}{c})=a-0.5\log(\tan(\Delta\mathrm{PA}_{i}))
\label{eq:v-pa}
\end{equation}
where $c$ is the speed of light, { and the  $\Delta\mathrm{PA}_\mathrm{i}$\ are the ones of Fig.  \ref{fig:PA}}. The slope is set to 0.5 to be consistent with  the assumption of Keplerian motion within the BLR. The intercept $a$ is a function of $M_{\rm BH}$ and can be written as
\begin{equation}
\label{eq:spmass}
a=0.5\log(\frac{GM_{{\rm BH}}\cos^{2}(\theta)}{c^{2}R_{{\rm sc}}})
\end{equation}
where $G$ is the gravitational constant, $R_{\rm sc}$ is the inner radius of the scattering region, and $\theta$ is the angle between the BLR and the scatterer plane. Based on our assumption of equatorial scattering, $\theta$ should be very close to 0. However, it is worth noting that a non-coplanar torus has $\theta \sim 10-20^\circ$, and thus may introduce a systematic underestimate on \mbh\ at $\sim 10\%$\ \citep{Afa2015}.

 { First, we verified that the slope is indeed consistent with a Keplerian velocity field. A bisector fit yields $b \approx -0.492 \pm   0.065$, consistent with the expected value $b = -0.5$.}  We then determined $a$ by fitting the velocity and PA by linear regression (shown in Fig.   \ref{fig:PA}). The estimated value is $a = -2.24 \pm 0.13$\ for \hb\ and $-2.22 \pm 0.15$ for \ha\  (with uncertainties at 1$\sigma$\ confidence level). 
To estimate the inner radius of the scattering region,  we apply two techniques. First, we consider the dust lag-luminosity correlation described in \citet{Koshida2014} that yields the following relation between the time lag in K band (days) and the luminosity in V band: 

\begin{equation}
\label{eq:rsc}
\log\text{\ensuremath{\Delta}}t_{\rm K}=a_{{\rm s}}+b_{\rm s}M_{{\rm V}}
\end{equation}
where $a_{\rm s} = -2.11 \pm 0.04.$ and $b_{\rm s} = -0.2$, both given by \citet{Koshida2014}. Fairall 9 is known as variable source \citep{santos-lleoetal97}, with secular trends superimposed to shorter time scale variation of a significant amplitude. We consider here the $V \approx$13.36 for the nucleus of Fairall 9 provided by \citet{griersmithvisvanathan79}, at an epoch when the AGN was reputed to be in a  ''bright'' phase.  The  corresponding  $V$ absolute magnitude is $M_{\rm V} = -23.0$, after applying a correction because of Galactic extinction ($A_\mathrm{V} \approx 0.071$). Here we assume that the time lag reflects the inner radius of the scattering region, i.e. $R_{{\rm sc}}=c\Delta t_{K} \approx 310 $ light days. Using Eq.  \ref{eq:spmass} we can estimate the black hole mass, which is shown in Table  \ref{tabSP}. 

The radius of the scattering region $R_\mathrm{sc}$ can also be estimated  following \citet{Afa2019} who define a relation between FUV observations and   $R_\mathrm{sc}$ from  GALEX observations. The specific at 1516 \AA\ for a 7 arcsec aperture is $\approx 3.08$ mJy. Trouble is, the FUV flux of Fairall 9 is strongly variable. The IUE average flux obtained during the reverberation mapping campaign of Fairall 9 by \citet{rodriguezpascualetal97} are consistent with the GALEX values, but the rms is $\approx$ 30\%, with the source doubling in flux on a timescale of 6 month. Applying the relation of  \citet{Afa2019} after extinction correction, we obtain $R_\mathrm{sc} \approx 215$ ld. { This second \mbh\ estimate is also reported on Table  \ref{tabSP}.}

\subsubsection{Other methods}

\paragraph{The RM method}
The RM  method is based on the response of the broad emission lines to the continuum variations. Reliable estimates of the lag between the continuum and the \hb\ broad emission line light curves for over 100 AGNs are presently available \citep[][and references therein]{du2019}.  The rest-frame time lag in \hb\ obtained using cross-correlation function (CCF) by \citet{peterson2004} is $\tau=17.4_{-4.3}^{+3.2}$ days. Alternatively, the rest-frame lag can be derived with a statistical technique  called JAVELIN \citep[][formerly known as SPEAR]{Zuetal2011}.  RM provides the line emitting region radius that can be entered in Eq.  \ref{eq3.2.2}.   { The measured line width for the full \hb\ profile is $\rm FWHM(\rm H\beta) = 4010 \pm 18$ \kms.}   The \hb\ lines has been studied in the past as the major virial broadening estimator for \mbh\ computation. 
In order to compute \mbh, we use velocity dispersion measures for the  BC
with corresponding virial factors, and with time lag obtained using different techniques. The final results for \mbh\ are reported in Table  \ref{tabRM}. 

The AGN Black Hole Mass Database \citep{bentzkatz15} reports the values of radii and line width measurements on the rms specta from the optical and UV reverberation mapping campaigns \citep{santos-lleoetal97,rodriguezpascualetal97}.  The last entry in Table  \ref{tabRM} refer to the \hb\ rms dispersion $\sigma$, with a $f \approx 3.76$.


\paragraph{The R-L scaling relations}
The RM observations of more than 70 AGNs so far have led to an empirical correlation between radius of the BLR and the luminosity of the continuum, i.e. the R-L relation \citep{Kaspi2000,bentz2009}. This relation has made possible  single-epoch \mbh\ estimations for large samples of quasars:  \mbh\ is derived just from the continuum (or emission line) luminosity  and line width of single-epoch spectroscopy \citep{vestergaard2002,VP2006,GreeneHo2005}. 
In Table  \ref{subtabRL}, the luminosity of the continuum at 5100 \AA\ is obtained from \citet{bentz2009}, which already has the host contamination subtracted. The nucleus luminosity of \hb\ BC is derived from \citet{PM2003}, and the luminosity of \ha\ is converted from $L_{5100}$ using the empirical relation given by \citet{GreeneHo2005}.


\paragraph{The $M_\mathrm{BH}-\sigma_{\star}$ relation}
The \mbh\ and the stellar velocity dispersion ($\sigma_{\star}$) in the host galaxy bulge has been found to be strongly correlated \citep{ferraresemerritt00,gebhardtetal00,nelson00}. However, for luminous Seyfert 1 galaxies like Fairall 9 it is relatively difficult to detect the virial motion in the host galaxy.  \citet{NW96} have found a strong correlation between $\sigma_{\star}$ and the FWHM of [O {\sc III}]$\lambda 5007$ for the majority of Seyfert galaxies, indicating that the [O {\sc III]}$\lambda 5007$ profile is dominated by virial motion in the bulge potential. Therefore we  use $\sigma_{\star}={\rm FWHM}_{\rm [O {\sc III}]}/2.35$  where ${\rm FWHM}_{\rm [O {\sc III}]}=425$ \kms\ { is measured by \citet{Whittle92} with the instrumental effects corrected}. The only direct stellar velocity dispersion measurement we are aware of: \citet{Oliva95} found $228 \pm 20$ \kms\ from the mid-IR feature of Si at $1.59\mu$m. The values are consistent but the IR value needs to be carefully calibrated, as a small difference can yield to a large mass difference. { In order to determine the correlation between \mbh\ and $\sigma_{\star}$,  previous work carried out by \citet{McConnell2011} separated the entire sample of 65 local galaxies into two sub-samples (early type and late type) based on their morphology.} 
We report in Table  \ref{msigma} the \mbh\ estimates    based on the early-type, late-type and entire sample scaling laws of \citet{McConnell2011}, although the scaling law derived for the entire sample should be preferred because of a better statistics.

{
\begin{table*}
\centering
\caption{\label{table2}\mbh\ Estimates}

    \subtable[\label{tabSP}Spectropolarimetry]{
    \setlength{\tabcolsep}{7pt}
    \begin{tabular}{cccccccc}
    \toprule 
      Line & $M_{V}$\tnote{1}    & $L_\mathrm{GALEX}(1516)$  & $R_{\rm sc}$ & $a$& $\cos{\theta}$ & \mbh\ & Notes/Refs. \tabularnewline
    & (mag) & (\ergs) & (ld) &  &  & ($10^{7}M_{{\rm \astrosun}}$)& \tabularnewline
    (1)  & (2)  & (3)& (4) &(5) &(6)& (7) &(8)
    \tabularnewline      
    \midrule 
        \hb\ &  \ldots &  2.75  & $215$    & $ -2.24 \pm 0.13 $& $\sim 1$ & $12.5 \pm 7.5$ &  \citet{Afa2019,feigelsonetal92}\tabularnewline 
         
      \hb\ & -23.0 & \ldots\  & $310$    & $ -2.24 \pm 0.13$ & $\sim 1$ & $18.2 \pm 10.8$ &  \citet{Afa2015,Koshida2014} 
      \tabularnewline
      \ha\ &  \ldots &  2.75  & $215$    & $ -2.22 \pm 0.15 $& $\sim 1$ & $13.7 \pm 9.5$ &  \citet{Afa2019,feigelsonetal92}\tabularnewline 
         
      \ha\ & -23.0 & \ldots\  & $310$    & $ -2.22 \pm 0.15$ & $\sim 1$ & $19.8 \pm 13.6$ &  \citet{Afa2015,Koshida2014} \tabularnewline
       \bottomrule
    \end{tabular}}
\begin{minipage}{\linewidth}
\vspace{0.2cm}
{ \textit{Notes}: \mbh\ estimated with spectropolarimetric data using Eq. \ref{eq:spmass}. Col. (1), emission line for mass estimation. Col. (2), absolute magnitude in $V$ band transformed from the nuclear apparent magnitude measured by \citet{griersmithvisvanathan79}. Col. (3), GALEX luminosity at 1516\AA. Col. (4), radius of the scattering region estimated with Eq. \ref{eq:rsc}(second and fourth rows) or  \citet{Afa2019} method (first and third rows). Col. (5), intercept of the linear regression as shown Fig. \ref{fig:PA}. Col. (6), cosine of the angle $\theta$ between BLR plane and the inner scattering region, which is assumed to be close to 1($\theta\sim0$). Col. (7), the estimated black hole mass with uncertainties at 1$\sigma$ confidential level. Col. (8), references for the apparent magnitude, GALEX luminosity, and their relation with $R_{\rm sc}$.}

\vspace{0.2cm}
\end{minipage}

\centering
\subtable[\label{tabRM}Virial - $r_\mathrm{BLR}$ from \hb\ RM]{
\setlength{\tabcolsep}{12pt}
\renewcommand{\arraystretch}{1.2}
\begin{tabular}{ccccccc}
\toprule 
Line component & $f$  & $R_{{\rm BLR}}$ & FWHM  & \mbh\ & Lag technique & Lag Ref \tabularnewline
  &  & (light days) & (${\rm km\,s^{-1}}$) & $(10^{7}M_{{\rm \astrosun}})$& &\tabularnewline
(1) & (2) & (3) & (4) & (5)& (6)& (7)\tabularnewline
\midrule

\hb\ BC & 2.18 & $17.4^{+3.2}_{-4.3}$ & $4010\pm18$  & $11.9^{+2.2}_{-2.9}$ &CCF & \citet{peterson2004} \tabularnewline
\hb\ BC & 2.18 & $19.4^{+42.1}_{-3.8}$ & $4010\pm18$  & $13.3^{+28.8}_{-2.6}$&JAVELIN &\citet{Zuetal2011} \tabularnewline
\hb\ rms $\sigma^\dagger$  & 3.76 & $17.4^{+3.2}_{-4.3}$ & $3787 \pm 197 $   &  $18.3^{+3.9}_{-4.9}$ & CCF & \citet{peterson2004} 
\tabularnewline
\bottomrule
\end{tabular}}
\begin{minipage}{\linewidth}
\vspace{0.2cm}
{ \textit{Notes:} $^\dagger$As reported in the AGN Black Hole Mass Database \citep{bentzkatz15}.\\
Col. (1), emission line and component. Col. (2), virial factor for corresponding line component(see Table \ref{tab:inc}. Col. (3), radius of the BLR obtained using different techniques. Col. (4), FWHM for different lines. The value for \ha\ has been rescaled to the correlating \hb\ width. Col. (5), black hole mass estimated with Eq. \ref{eq3.2.2}. Col. (6), techniques for obtaining the time lags. Col. (7), references for the lags.}
\vspace{0.2cm}
\end{minipage}

\centering
\subtable[\label{subtabRL}Scaling laws]{
\setlength{\tabcolsep}{15pt}
\begin{tabular}{ccccc}
\toprule 
Relation & FWHM & $L$  & \mbh\ & Relation Ref \tabularnewline
 & (\kms)  & (\ergs) & $(10^{7}M_{{\rm \astrosun}})$ &\tabularnewline
(1) & (2) & (3) & (4) & (5)  \tabularnewline
\midrule
FWHM$_{\rm H\beta}$, $L_{5100}$&  $4010 \pm 18 $  &$1.38 \pm 0.08$ & {$15.4 \pm 0.8$} & \citet{VP2006}\tabularnewline
FWHM$_{\rm H\alpha}$, $L_{5100}$&  $3847.8 \pm 9.6$ &$1.38 \pm 0.08$ & $24.1 \pm 1.4$ & \citet{bentz2009,Assef2011}
\tabularnewline
FWHM$_{\rm H\alpha}$, $L_{\rm H\alpha}$&  $3847.8 \pm 9.6$ & $0.15\pm0.01$ & $31.7 \pm 6.5$ & \citet{GreeneHo2005,Shenetal2011} \tabularnewline
\bottomrule
\end{tabular}}
\begin{minipage}{\linewidth}
\vspace{0.2cm}
{ \textit{Notes:} single epoch \mbh\ estimated with scaling relations. Col. (1), empirical relations between the emission line FWHM and the luminosity. Col. (2), emission line FWHM. Col. (3), luminosity of the continuum or emission line. Col. (4), the estimated \mbh. Col. (5), references for the luminosity and the scaling relation.}
\vspace{0.2cm}
\end{minipage}

\centering
\subtable[\label{msigma} \mbh$-\sigma_\star$ correlation]{
\setlength{\tabcolsep}{15pt}
\begin{tabular}{ccccc}
\toprule 
 Samples &  $\sigma_{\star}$& Line & \mbh\ & $\sigma_{\star}$ Ref \tabularnewline
 & (\kms) & &($10^{7}$\msun)&  \tabularnewline
(1) & (2) & (3)& (4) &(5)\tabularnewline
\midrule
 entire & $215 \pm 20$ & Si(1.59$\mu$m) & {$28.2 \pm 15.5$}&\citet{Oliva95} \tabularnewline
  entire & $181\pm10 $ & [O {\sc iii}]$\lambda5007$ & {$11.7 \pm 3.7$}&\citet{Whittle92} \tabularnewline
 early type & $215 \pm 20$ & Si(1.59$\mu$m) &  {$33.3 \pm 16.1 $}&\citet{Oliva95} \tabularnewline
 early type & $181\pm10$& [O {\sc iii}]$\lambda5007$ &  {$15.2 \pm 4.4 $}&\citet{Whittle92} \tabularnewline

\bottomrule\addlinespace[1ex]
\end{tabular}}

\begin{minipage}{\linewidth}
\vspace{0.2cm}
{ \textit{Notes:} \mbh\ estimated with the empirical \mbh$-\sigma_\star$ correlation. The  empirical \mbh$-\sigma_\star$ correlation of \citet{McConnell2011} is provided for three different samples analyzed in their work. Col. (1), samples for regression in \citet{McConnell2011}. Col. (2), stellar dispersion. Col. (3), emission/absorption line for stellar dispersion estimation. Col. (4), the estimated \mbh. Col. (5), references of the  emission/absorption line dispersion. }
\vspace{0.2cm}
\end{minipage}

\end{table*}

}



\vfill\clearpage
\pagebreak

\section{Polarization modeling}
\label{models}

 \begin{figure*}
    \centering
    {\includegraphics[width=0.8\textwidth]{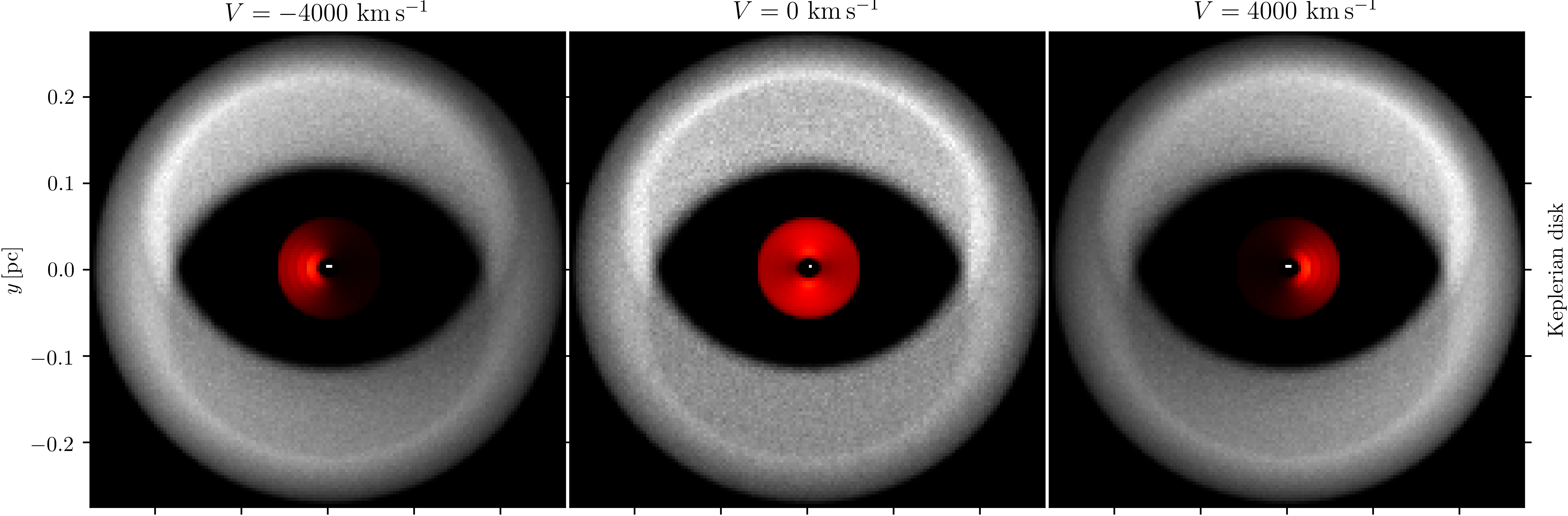}}
    {\includegraphics[width=0.8\textwidth]{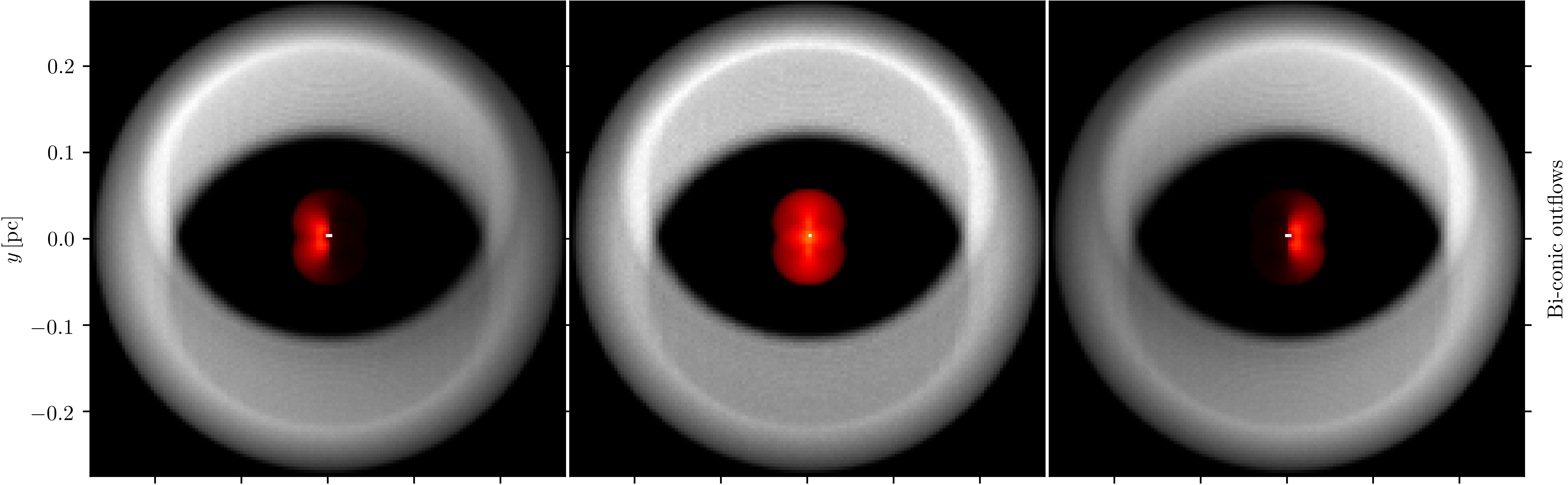}}
    {\includegraphics[width=0.8\textwidth]{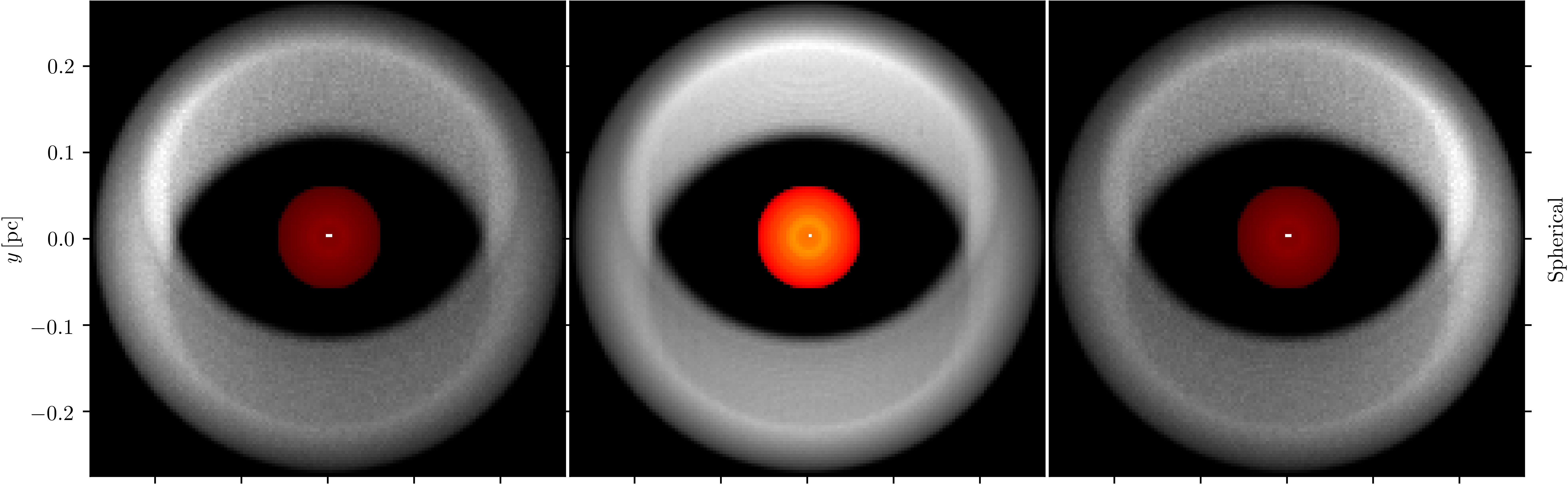}} 
    {\includegraphics[width=0.8\textwidth]{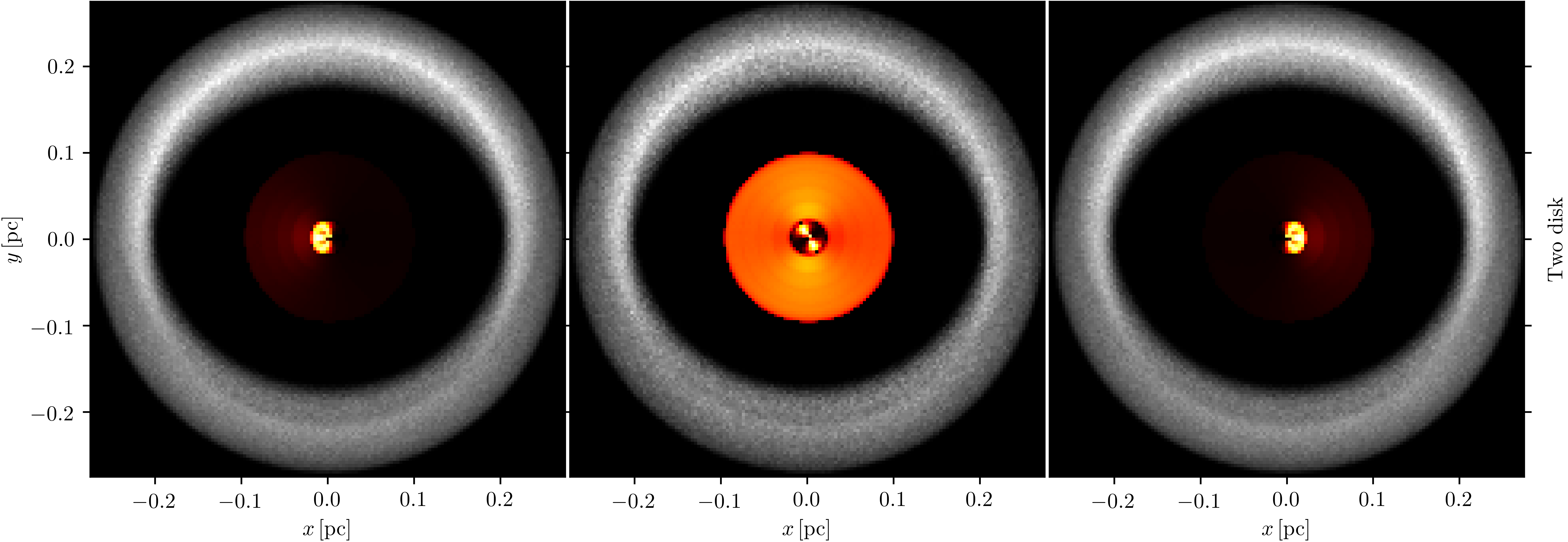}}
    
    \caption{Geometric setup of our models. BLR geometry from top to bottom: Keplerian disk, double-cone, spherical and two disk model. From left to right: radiation emitted at V = -4000, 0 and 4000 \kms. The scattering region is shown in gray. The point source approximates the accretion disk emission. All the models except the two disk model are viewed at an inclination of $25^\circ$. The two disk model is viewed at an inclination of the outer disk $15^\circ$ and azimuth $\phi = 45^\circ$.
    }
    \label{fig:model_geometry}
\end{figure*}


\begin{figure*}
    \centering
    \includegraphics[width=0.9\hsize]{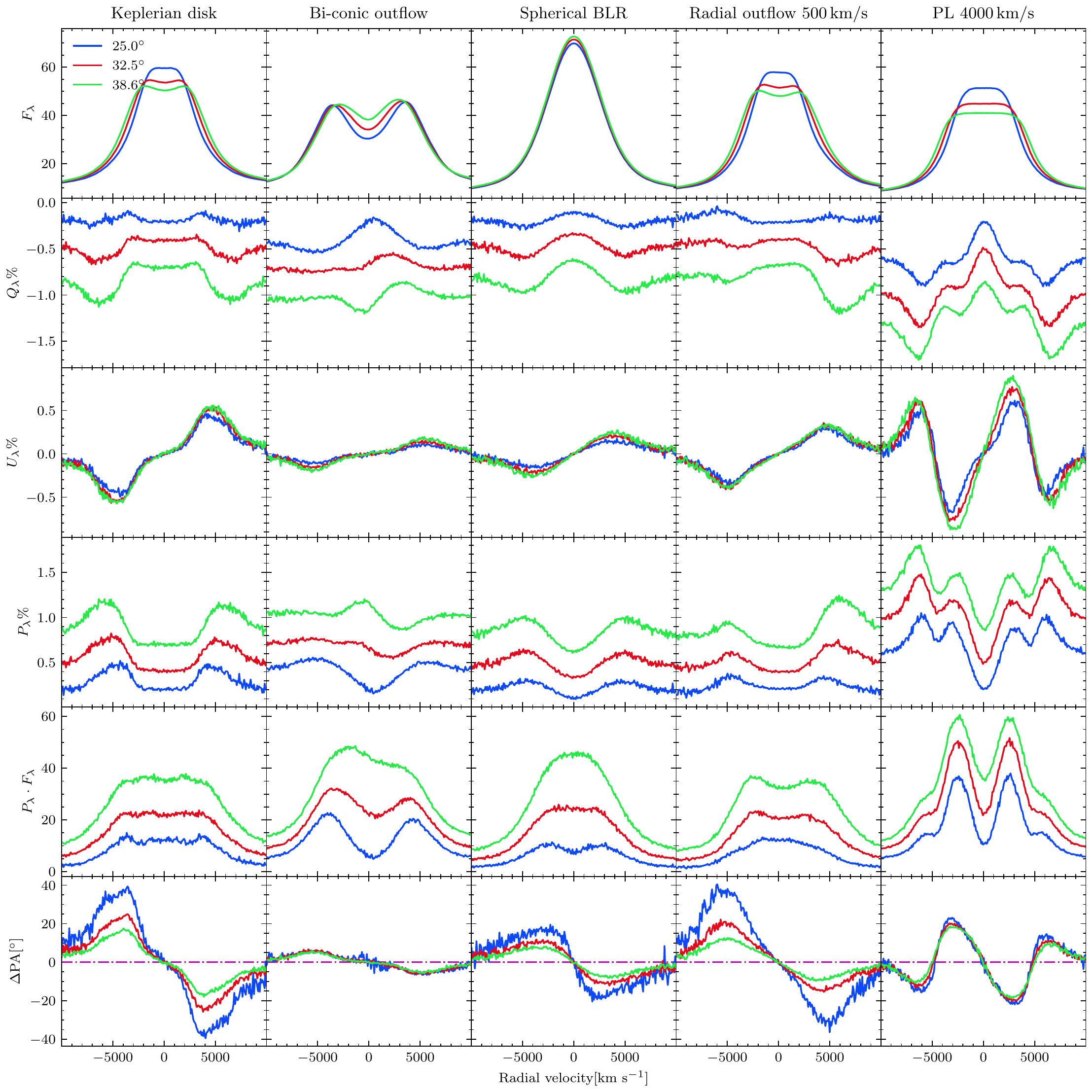}
    \caption{Polarization parameters for  three of the models (Keplerian disk, biconical outflow, and spherical BLR) considered in this work, at three different inclination values: total unpolarized flux (top row), polarized flux (second from top), Stokes parameters $Q_\lambda$\ and $U_\lambda$, degree of linear polarization $P_\lambda$ (second row from bottom), polarization position angle(bottom).} \label{fig:Fairall9_STO_SED.pdf}
\end{figure*}

\begin{figure*}
    \centering
    \includegraphics[width=0.7\hsize,keepaspectratio = true]{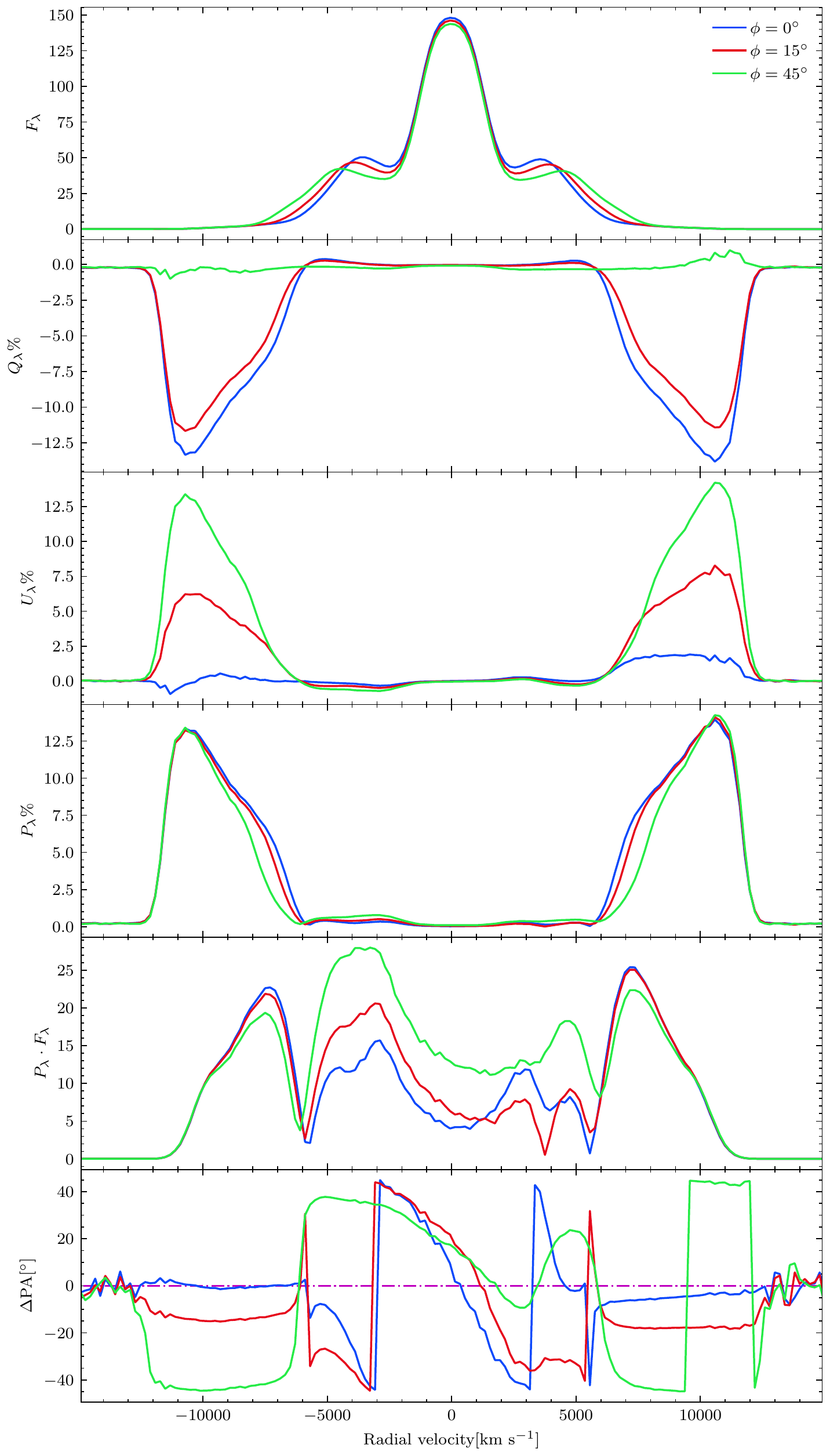}
    \caption{Same as Fig.\, \ref{fig:Fairall9_STO_SED.pdf}, but for the two disk model. The system is viewed at inclination $i = 15^\circ$ and azimuth $\phi = 0^\circ,\,15^\circ$and $45^\circ$ (blue, red and green lines respectively).}
    \label{fig:Fairall9_14_SED.pdf}
    \label{lastpage}
\end{figure*}

In order to investigate various effects of different BLR geometry and kinematics on the polarization spectra, we performed three simulations using radiative transfer codes \textsc{stokes}\footnote{\url{http://www.stokes-program.info}} \citep{GG07,Marin12,Marin15,Rojas18,Marin18}\ and \textsc{skirt}\footnote{\url{https://skirt.ugent.be/root/_landing.html}} \citep[][and references therein]{Baes19,Camps20}. We assumed that the BLR emission is being scattered by the inner part of the dusty torus. We assumed three different geometries of the BLR: disk-like, bi-conic and spherical (Fig.\, \ref{fig:model_geometry}).

The velocity field for each BLR geometry was chosen respectively as Keplerian, radial outflows and random motion. The inner radius of the BLR was taken from the RM measurements as the mean value: $R_{\mathrm{in}}^{\mathrm{BLR}} = 18\,\mathrm{ld}$. The BLR outer radius was calculated based on bolometric luminosity $R_{\mathrm{out}}^{\mathrm{BLR}} = 0.2L_{46}^{0.5}\,\mathrm{pc} = 70\,\mathrm{ld}$\ \citep{DL93}. The bolometric luminosity was calculated from optical luminosity $L_{5100}$\ \citep{Runnoe12}. The inner radius of the scattering region $R_{\mathrm{in}}^{\mathrm{SR}} = 262\,\mathrm{ld}$\ was estimated from the UV $L_{\textsc{galex}}(1516)$\ \citep{Afa2019}. The outer radius of the SR was taken as the distance in the equatorial plane from which the half-opening angle of the outer edge of the BLR is viewed at the angle of $25^\circ$, i.e.\,$R_\mathrm{out}^{\mathrm{SR}} = R_{\mathrm{out}}^{\mathrm{BLR}}/\sin 12.5^\circ = 322\,\mathrm{ld}$\ \citep[same as it was done by][]{Savic2018}. The total radial depth of the SR is set to unity. The SMBH mass was set to $M_{\mathrm{BH}} = 2\times10^{8}\,M_\odot$.  All parameters of the model using the mean value of each observable given in Table  \ref{table2}. For every simulation, we used one BLR configuration while keeping the same SR. Each of the BLR configuration has the same inner and outer radius.

In Fig.\, \ref{fig:Fairall9_STO_SED.pdf} (due to azimuthal symmetry), we show the profiles of the total unpolarized flux (top panels), polarized flux $p\times F$ (top second panels), Stokes parameters $Q$ (top third panel), $U$ (top fourth panels), the degree of linear polarization $p$ (top fifth panels) and PA(bottom panels) for the disk-like, bi-conic and spherical BLR geometry (from left to right respectively). 

A warped disk was modeled as two disks: the inner and the outer. The inner disk was set from 1000 $r_{\rm g}$ to 2000 $r_{\rm g}$ and rotated along the $x-$axis for $30^\circ$. The outer disk continues from 2000 $r_{\rm g}$ up to 10000 $r_{\rm g}$ Fig.\, \ref{fig:model_geometry} (bottom panels). The system is viewed at an inclination $i = 15^\circ$ and at azimuthal angles $\phi = 0^\circ,\,\,22.5^\circ\,\,45^\circ$ and $90^\circ$.
Polarization profiles for the two disk model is shown in Fig.\, \ref{fig:Fairall9_14_SED.pdf}.

\end{document}